\documentclass[12pt]{article}
\usepackage{epsfig}
\usepackage{cite}

\begin{document}
\begin{titlepage}

\centerline{\large\bf Neutron electric polarizability from unquenched
lattice QCD}
\vspace{0.2cm}

\centerline{\large\bf using the background field approach}

\bigskip
\centerline{M.~Engelhardt}
\vspace{0.2cm}
\centerline{(LHPC Collaboration)}
\vspace{0.5cm}
\centerline{\em Physics Department, New Mexico State University}
\centerline{\em Las Cruces, NM 88003, USA}
\vspace{0.5cm}

\abstract{A calculational scheme for obtaining the electric polarizability
of the neutron in lattice QCD with dynamical quarks is developed, using
the background field approach. The scheme differs substantially from
methods previously used in the quenched approximation, the physical
reason being that the QCD ensemble is no longer independent of the
external electromagnetic field in the dynamical quark case. One is
led to compute (certain integrals over) four-point functions.
Particular emphasis is also placed on the physical role of constant
external gauge fields on a finite lattice; the presence of these fields
complicates the extraction of polarizabilities, since it gives rise to
an additional shift of the neutron mass unrelated to polarizability
effects. The method is tested on a $SU(3)$ flavor-symmetric ensemble
furnished by the MILC Collaboration, corresponding to a pion mass of
$m_{\pi } =759\, \mbox{MeV} $. Disconnected diagrams are evaluated using
stochastic estimation. A small negative electric polarizability of
$\alpha =(-2.0\pm 0.9)\cdot 10^{-4} \, \mbox{fm}^{3} $ is found for
the neutron at this rather large pion mass; this result does not
seem implausible in view of the qualitative behavior of $\alpha $
as a function of $m_{\pi } $ suggested by Chiral Effective Theory.}

\vspace{1cm}



\end{titlepage}

\section{Introduction}
\label{intsec}
An important characteristic of hadrons is their stiffness when subjected
to outside forces, which are typically conveyed by external electromagnetic
fields. This response is summarized in hadron polarizabilities.
Understanding these quantities will contribute to making hadron
structure more palpable. Experimentally, polarizabilities are accessible,
e.g., via soft Compton scattering; heuristically, in such an experiment,
the photon electric and magnetic fields polarize the target hadron, which
in turn manifests itself in the Compton scattering amplitude observed.
Accordingly, polarizabilities are effects of second order in the external
fields.

The aforementioned sensitivity of low-energy Compton scattering to hadron
structure can be cast in precise terminology \cite{polarization}, permitting
stringent tests of theoretical understanding of that structure. Starting
with the leading order in the low-energy expansion, the non-Born
(i.e., structure-dependent) part of the scattering amplitude is determined
by the static dipole electric and magnetic polarizabilities $\alpha $
and $\beta $. These are given by the hadron mass shift in the
presence of external static electric and magnetic fields, specifically
the part of the mass shift which depends quadratically on those fields,
in accordance with a (spin-independent) effective dipole interaction
Hamiltonian
\begin{equation}
H_{eff}^{(2)} = -\frac{1}{2} \left( \alpha E^2 + \beta B^2 \right) \ .
\label{leadpol}
\end{equation}
The present investigation focuses on the electric polarizability $\alpha $
of the neutron.

Lattice hadron polarizability calculations have hitherto been carried
out only in the quenched approximation
\cite{fiebpol,wilann,wildmo,wile,wilb,wilepap,wilbpap,wildub}.
The reason for this lies in the fact that, in the case of polarizabilities,
the complication implied by going from a quenched to an unquenched
calculation involves more than just the usual vastly increased effort
required to generate a dynamical quark ensemble. In addition, a quenched
calculation is simpler due to the gauge ensemble being independent
of the external electromagnetic field; after all, the only way the external
field can influence the gauge ensemble is through the quarks, whose
backreaction on the gauge fields is precisely truncated in a quenched
calculation. This is no longer true in the dynamical quark case.

This physical difference manifests itself formally in the fact that
substantially different computational schemes have to be used in the
dynamical quark case as compared to the quenched case. In the quenched
case, one can simply generate gauge configurations in the absence of the
external electromagnetic field and introduce the latter a posteriori by
an appropriate modification of the link variables in those configurations.
The requisite hadron two-point functions are then evaluated directly using
the modified gauge configurations. By contrast, in a fully dynamical
calculation, as discussed in more detail below, one in principle would
need to generate the gauge ensemble anew for each external field considered.
The prohibitive cost of such a scheme can be mitigated to some extent by
expanding in the external field, leading, in effect, to the calculation
of (certain space-time integrals over) four-point functions. In general,
these include disconnected contributions. However, even resorting to
such a four-point function method leads to a substantially more expensive
calculation than one is confronted with in the quenched case; an early
exploratory study of four-point function methods \cite{wilann} (using
a quenched ensemble) highlights this point\footnote{The four-point
functions considered in \cite{wilann} are different from the ones which
are calculated in the present work, since different theoretical approaches
are used. However, the computational complexity resulting from the two
approaches is similar.}.

The most of this situation has been made hitherto in a series of
investigations \cite{fiebpol,wile,wilb,wilepap,wilbpap,wildub}
taking full advantage of the simplifications offered by the quenched
approximation. An initial study of the electric polarizability
of neutral hadrons \cite{fiebpol} using staggered fermions yielded
results both for the neutral pion and the neutron. This was later
extended to include all neutral members of the baryon octet, as well
as the baryon decuplet and the vector meson octet, in investigations
using both Wilson and clover fermions \cite{wile,wilepap}; these
studies surveyed a range of pion masses down to about $500\, \mbox{MeV} $.
Also the magnetic polarizability of a wide array of hadrons was
investigated using the same range of pion masses and fermion actions
\cite{wilb,wilbpap,wildub}; these studies included also charged
hadrons, in particular the entire baryon octet and decuplet as well as
selected pseudoscalar and vector mesons.

The present work extends the aforementioned work in two main aspects:
\begin{itemize}
\item Use of a dynamical quark ensemble: As discussed above, dynamical
quark calculations of polarizabilities were rendered intractable in the
past by the associated computational cost. Recent increases in available
computing resources are making quantitative four-point function calculations,
appropriate for dynamical ensembles, feasible. This investigation presents
the first result for the electric polarizability of the neutron in a
dynamical quark ensemble, albeit obtained at a still rather heavy pion
mass of $759\, \mbox{MeV} $.
\item Recognition that, on a finite spatial volume, a constant gauge
field is not a pure gauge, but has physical consequences which must
be disentangled from polarizability effects.
\end{itemize}
Some elaboration on the latter issue, which is also relevant in the
quenched approximation, is useful at this point. As noted further
above, hadron polarizabilities can be probed via the mass shift in the
presence of external electromagnetic fields. The hitherto preferred
method \cite{fiebpol,wile,wilepap} of introducing a constant electric
field in, say, the 3-direction is to represent it by a non-vanishing
3-component of the gauge field,
\begin{equation}
A_3 = E(t-t_0 ) \ .
\label{a3def}
\end{equation}
This choice has the advantage that jumps in the gauge field at the lattice
boundaries (inducing spurious localized electric fields there) occur only
in the temporal direction. In this case, hadron two-point functions
evaluated in the bulk of the lattice are insensitive to the jumps, which
only occur far in the past or the future from the point of view of the
measurement.

However, there is an ambiguity in the prescription (\ref{a3def}), namely,
at which time $t_0 $ one chooses to begin counting time. Different choices
of $t_0 $ correspond to different constant shifts of $A_3 $. Working in a
spatially infinite setting, this ambiguity would be inconsequential, since
constant gauge fields are then pure gauges. However, on a finite space, the
spatial boundary conditions (which in the following will be taken to be
periodic) restrict the available gauge transformations and only allow for
discrete shifts of the gauge fields. As a simple example, consider a charged
particle in a constant field on a circle of length $L$ described by the
Hamiltonian $H=(-i\partial_{x} +A)^2 $. Its energy eigenvalues are
$E_n = (2\pi n/L +A)^2 $, where $n$ can be any integer. The ground state
energy therefore is $E_0 = A^2 $ as long as $A\in [-\pi /L , \pi /L]$, and
$E_0 $ is periodic in $A$ with period $2\pi /L$, reflecting the residual
discrete gauge invariance. The spectrum explicitly depends on $A$ and,
compared with the case $A=0$, the ground state energy can deviate by as
much as $\Delta E_0 = \pi^{2} /L^2 $. Thus, while this is ultimately
nothing but a finite size effect, it vanishes rather slowly (only as a
power of $L$) as the spatial volume is increased. It is a priori unclear
how difficult it is in practice to deal with this effect by using different
lattice sizes. Within the present investigation, that avenue is closed at
any rate, since the dynamical quark ensemble which will be used is only
available at one spatial volume. Instead, measurements at several different
$t_0 $ in (\ref{a3def}) will be used in order to treat this effect.

Another aspect of the same issue is that the Hamiltonian in the presence of
the field (\ref{a3def}) is not time-independent. Physics at two widely
separated times differ precisely by a shift in the external gauge field
$A_3 $. If the electric field $E$ is very small, the strong dynamics can
instantaneously adjust to the change in the external field as time passes;
one will observe an adiabatic change in the physical spectrum. The hadronic
two-point function will generally not fall off as a simple exponential in
Euclidean time, even for large such times. Both the hadron's energy as
well as its wave function will contain time dependences. This is
reminiscent of the behavior induced by the acceleration of charged hadrons
in the electric field \cite{detmold}. That particular effect is negligible
for sufficiently heavy hadrons, such as nucleons. By contrast, the constant
gauge field effect discussed here is one of the dominant effects, and care
needs to be taken to disentangle it from the nucleon polarizability.

A way to avoid the time dependence discussed above is to instead use a
gauge field representation of the type
\begin{equation}
A_0 = -Ex_3
\label{a0def}
\end{equation}
inducing the same external electric field as (\ref{a3def}); indeed, in the
present work, also this case will be investigated. However, it should be
noted that this choice also has disadvantages which, on balance, are
no less problematic than the time dependence engendered by (\ref{a3def}).
Namely, the advantage of time-independence using (\ref{a0def}) is offset
by the fact that spatial translational invariance is lost; the representation
(\ref{a0def}) conflicts with spatial periodicity and the periodic boundary
conditions enforce a spike in the electric field resulting from the jump
in $A_0 $ as one crosses the boundary of the lattice in the
3-direction\footnote{Note that, in the lattice formulation, there
exist discrete choices of $E$ which mitigate this problem, namely,
integer multiples of $2\pi /aL$, where $L$ is the extent of the lattice
in the relevant direction and $a$ denotes the lattice spacing. However,
this argument relies on the compactness of the gauge link variables and,
on realistic lattices, corresponds to strong electric fields. By contrast,
hadron electric polarizabilities are given specifically by the term
quadratic in $E$ of a Taylor expansion of their mass. To isolate this
term, it is necessary to vary $E$ over a denser set of values than
provided by the aforementioned discrete choices, for which the Taylor
expansion will generally not converge well on lattices of a practical
size.}. In effect, the neutron thus propagates in a spatially varying
potential and its energy contains, e.g., contributions from quantum
mechanical zero-point motion in that potential. Also the neutron's
internal wave function is distorted by the electric field spikes.
As a consequence, it is not straightforward to isolate the polarizability
from the full measured mass shift.

\section{Measurement method}
\subsection{Neutron two-point function}
\label{2ptsec}
The objective of the present investigation is to extract the neutron
mass from the neutron two-point function in the presence of an external
electric field. The neutron two-point function is the correlator
\begin{equation}
\langle N_{\alpha^{\prime } } (x^{\prime } ) \bar{N}_{\alpha } (x) \rangle
=\frac{1}{Z} \int [DU] [D\bar{\psi } ] [D\psi ] 
\exp (-S[\psi , \bar{\psi } , U])
N_{\alpha^{\prime } } (x^{\prime } ) \bar{N}_{\alpha } (x) \ ,
\label{2ptdef}
\end{equation}
with the lattice discretization of the functional integral to be specified
below. Both the action $S$ and the (smeared) neutron fields $N$, $\bar{N} $
in general depend on the external electromagnetic field $A_{\mu } $. At
face value, this would imply that one needs to generate lattice ensembles
using an action modified by the external field in order to evaluate
(\ref{2ptdef}). That would clearly be prohibitively expensive using
dynamical quarks. However, decomposing the action as
\begin{equation}
S=S_0 +S_E \ ,
\end{equation}
where $S_0 $ is the action in the case of vanishing external field, one can
rewrite (\ref{2ptdef}) as
\begin{equation}
\langle N_{\alpha^{\prime } } (x^{\prime } ) \bar{N}_{\alpha } (x) \rangle
=\frac{\langle e^{-S_E } N_{\alpha^{\prime } } (x^{\prime } )
\bar{N}_{\alpha } (x) \rangle_{0} }{\langle e^{-S_E } \rangle_{0} } \ ,
\label{2ptfrac}
\end{equation}
with $\langle \ldots \rangle_{0} $ denoting the average in the absence of
the external field,
\begin{equation}
\langle O \rangle_{0} = \frac{1}{Z_{0} } \int [DU] [D\bar{\psi } ] [D\psi ]
\exp (-S_0 ) O \ .
\label{expect0}
\end{equation}
While this reduces the problem to integrations over the lattice ensemble
in the absence of the external field, expectation values such as the ones
in (\ref{2ptfrac}) generally suffer from severe overlap problems. The
crucial step which renders the problem somewhat more tractable results
from the fact that it is sufficient to know the quadratic term in the
Taylor expansion of (\ref{2ptfrac}) with respect to the external field
in order to extract the neutron polarizability. Then, one can expand
\begin{equation}
\exp (-S_E ) = 1-S_E +S_E^2 /2 +\ldots
\end{equation}
and the evaluation of (\ref{2ptfrac}) reduces to the calculation of
certain space-time integrals over four-point functions, as will be
discussed in detail below. Before proceeding to describe this perturbative
expansion, it is now possible to specify how the functional integration
in (\ref{expect0}) will be carried out. As usual, decomposing $S_0 $
into its pure gauge and fermion parts,
\begin{equation}
S_0 = S_G + S_F \ ,
\end{equation}
the integration is cast in terms of an average over an ensemble of gauge
fields $U$,
\begin{equation}
\langle O \rangle_{0} = \frac{1}{Z_{0} } \int [DU]
\exp (-S_{G,eff} ) \langle O \rangle_{U} \ ,
\label{milcav}
\end{equation}
governed by the action $S_{G,eff} $ which includes the effects of both the
pure gauge term as well as the determinant of the Dirac operator from
$S_F $. Adopting $2+1$ flavor Asqtad quark fields to evaluate the
determinant, one can utilize the corresponding dynamical quark ensembles
made available by the MILC Collaboration \cite{milc1,milc2,milc3}. The
numerical results reported in the present work were obtained using $99$
configurations from the $SU(3)$ flavor-symmetric ensemble with quark
masses given by $am_s =am_l =0.05$, where the lattice spacing
$a=0.124\, \mbox{fm} $ is determined by heavy quark spectroscopy
\cite{aubin}. Computation at such a relatively large quark mass is
comparatively inexpensive and serves primarily to validate the concepts
developed in this work as well as giving a first indication of the
feasibility of a broader calculational effort within the framework
advanced here.

The aforementioned configurations were originally generated on
$20^{3} \times 64$ lattices. In the present work, these lattices were
chopped in half in the time direction, i.e., measurements were carried
out on $20^{3} \times 32$ lattices containing the first $32$ time slices
of the original $20^{3} \times 64$ lattices. Correspondingly, quark
propagators determining the quantities $\langle O \rangle_{U} $,
cf.~(\ref{ou}), were evaluated using Dirichlet boundary conditions
at the temporal edges of the chopped lattices. The lattices were furthermore
HYP-smeared \cite{hypsmear} to reduce the effect of dislocations.

For a given gauge configuration $U$, the expectation value
\begin{equation}
\langle O \rangle_{U} = \frac{\int [D\bar{\psi } ] [D\psi ]
\exp (-S_F ) O}{\int [D\bar{\psi } ] [D\psi ] \exp (-S_F )}
\label{ou}
\end{equation}
needs to be evaluated. At this point, a hybrid approach is adopted
\cite{gapaper,gpdpaper}: While the gauge ensemble used in the average
(\ref{milcav}) is generated using Asqtad quarks, (\ref{ou}) is evaluated
using domain-wall quarks \cite{kaplan,shamir}. The reason for this choice
lies in the longer-term goal of extending the present investigation to
light quark masses at which a chirally well-behaved quark discretization
becomes important. Thus, the fields $\psi $, $\bar{\psi } $ in (\ref{ou})
are taken to be five-dimensional, $\psi (x,s)$ and $\bar{\psi } (x,s)$,
where $x$ labels four-dimensional space-time and $s$ labels the fifth
coordinate. The latter is subdivided into $L_s =16$ spacings,
$s\in \{ 0,\ldots ,L_s -1 \} $, and $\psi (x,s) \equiv 0$ outside that
range (i.e., derivatives in the fifth direction have hard boundaries).
This value of $L_s $ is sufficient to keep the residual mass $m_{res} $
which characterizes explicit chiral symmetry breaking by the domain-wall
fermion discretization suppressed by more than an order of magnitude
compared to the quark mass discussed further below \cite{gapaper,gpdpaper}.
The boundaries $s=0$ and $s=L_s -1$ provide domain walls which support
quasi--four-dimensional light physical quark modes; left-handed modes
are bound to $s=0$ and right-handed modes to $s=L_s -1$. It is useful
to define corresponding four-dimensional projected quark fields
\begin{eqnarray}
\Psi (x) &=& \frac{1-\gamma_{5} }{2} \psi (x,0)
+ \frac{1+\gamma_{5} }{2} \psi (x,L_s -1) \\
\bar{\Psi } (x) &=& \bar{\psi } (x,0) \frac{1+\gamma_{5} }{2}
+ \bar{\psi } (x,L_s -1) \frac{1-\gamma_{5} }{2} \ .
\label{projq4}
\end{eqnarray}
In terms of the above fields, the action $S_F $ in (\ref{ou}) reads
\begin{eqnarray}
S_F [\psi , \bar{\psi } , U] &=& -\sum_{x,s} \sum_{\mu } \bar{\psi } (x,s)
\left( \frac{1-\gamma_{\mu } }{2} \left[ U_{\mu } (x) \psi (x+e_{\mu }, s)
- \psi (x,s) \right] \right. \nonumber \\
& & \ \ \ \ \ \ \ \left.
-\frac{1+\gamma_{\mu } }{2} \left[ -U^{\dagger }_{\mu } (x-e_{\mu } )
\psi (x-e_{\mu }, s) + \psi (x,s) \right] \right) \nonumber \\
& & -\sum_{x,s} \bar{\psi } (x,s) M_5 \psi (x,s)
+\sum_{x} \bar{\Psi } (x) m_f \Psi (x) \ ,
\label{sdwf}
\end{eqnarray}
where $\mu $ runs over all five dimensions and $U_5 \equiv 1$. Note
that the fermion fields also carry a flavor index; in the final term,
i.e., the quark mass term, which is constructed directly in terms of
the projected four-dimensional quark fields $\Psi $ and $\bar{\Psi } $,
$m_f $ in general represents a (diagonal) matrix in flavor space. In
the $SU(3)$ flavor-symmetric case studied here, $m_f $ is given by
one single number.

Using spectral flow analyses, the five-dimensional mass parameter $M_5 $
in (\ref{sdwf}) was chosen to take the value $M_5 =1.7$ in order to
optimize the chiral properties \cite{gapaper,gpdpaper}. Finally, the
quark mass was adjusted such as to match the pion mass obtained in the
present hybrid approach to the lightest pion mass extracted from a
pure Asqtad calculation \cite{aubin}; this yields \cite{gapaper,gpdpaper}
the choice $am_f =0.081$.

The domain wall fermion action (\ref{sdwf}) also determines the interaction
between the quarks and the external electric field. The additional
electromagnetic gauge field $A_{\mu } $ generating the external
electric field modifies the gauge link variables,
\begin{equation}
U_{\mu } (x) \ \longrightarrow \ \exp (iaq_f A_{\mu } (x) ) U_{\mu } (x) \ ,
\label{amuintro}
\end{equation}
where $a$ denotes the lattice spacing; note that the fractional electric
charge $q_f $ varies according to flavor. Note also that the particular
forms of $A_{\mu } $ used in this work, cf.~(\ref{a3def}),(\ref{a0def}),
are all such that $A_{\mu } $ is constant in the $\mu $-direction; hence
the simple form (\ref{amuintro}) for the exponentiated integral along the
link. Inserting the modified link variables (\ref{amuintro}) into the
domain wall fermion action (\ref{sdwf}) and separating off the part which
remains for vanishing external field, $A_{\mu } =0$, yields the
five-dimensional interaction
\begin{eqnarray}
S_{E,5d} &=& -\sum_{x,s} \sum_{\mu } \bar{\psi } (x,s) \left(
\frac{1-\gamma_{\mu } }{2} \left( e^{iaq_f A_{\mu } (x) } -1 \right)
U_{\mu } (x) \psi (x+e_{\mu }, s) \right. \label{5dcoup} \\
& & \ \ \ \ \ \ \ \ \ \ \ \ \ \ \ \ \
+\left. \frac{1+\gamma_{\mu } }{2}
\left( e^{-iaq_f A_{\mu } (x-e_{\mu } ) } -1 \right)
U^{\dagger }_{\mu } (x-e_{\mu } ) \psi (x-e_{\mu }, s) \right) \nonumber
\end{eqnarray}
generating a vertex which couples the five-dimensional domain wall fermion
fields $\psi $, $\bar{\psi } $ to the external field.

To arrive at a practicable computational scheme, in the calculations
presented further below, the external gauge field $A_{\mu } $ is not
coupled directly to the five-dimensional fields according to
(\ref{5dcoup}), but instead to the corresponding four-dimensional
projected quark fields $\Psi $, $\bar{\Psi } $. Accordingly, a
renormalization factor $z_V $ must be included with the four-dimensional
coupling to compensate for the effect of the projection of the quark
fields. Thus, the modified interaction vertex used in practice is
\begin{eqnarray}
S_E &=& -z_V \sum_{x} \sum_{\mu } \bar{\Psi } (x) \left(
\frac{1-\gamma_{\mu } }{2} \left( e^{iaq_f A_{\mu } (x) } -1 \right)
U_{\mu } (x) \Psi (x+e_{\mu } ) \right. \label{4dcoup} \\
& & \ \ \ \ \ \ \ \ \ \ \ \ \ \ \ \ \
+\left. \frac{1+\gamma_{\mu } }{2}
\left( e^{-iaq_f A_{\mu } (x-e_{\mu } ) } -1 \right)
U^{\dagger }_{\mu } (x-e_{\mu } ) \Psi (x-e_{\mu } ) \right) \ . \nonumber
\end{eqnarray}
The renormalization factor $z_V $ will be determined in section
\ref{rensec}. The reason for the adoption of the modified interaction
(\ref{4dcoup}) lies in the practical expense of storing full
five-dimensional propagators as opposed to ones which have been
projected down to four dimensions at source and sink. This modus
operandi constitutes a compromise which certainly should be revisited
as storage constraints change. Using full five-dimensional propagators
and coupling the conserved five-dimensional current to the external
electromagnetic field directly via (\ref{5dcoup}) would be the most
consistent treatment, and would eliminate the need for renormalization
of the interaction vertex.

Finally, it is necessary to specify the neutron sources and sinks
$\bar{N} $, $N$ in (\ref{2ptdef}):
\begin{eqnarray}
N_{\alpha } (x) &=& \delta_{\alpha \beta } (C\gamma_{5} )_{\gamma \delta }
\epsilon_{bcd} Q_{b\beta }^{(d)} (x) Q_{c\gamma }^{(d)} (x)
Q_{d\delta }^{(u)} (x)
\label{ndef} \\
\bar{N}_{\alpha } (x) &=& \bar{Q}_{d\delta }^{(u)} (x)
\bar{Q}_{c\gamma }^{(d)} (x) \bar{Q}_{b\beta }^{(d)} (x)
\epsilon_{bcd} \delta_{\alpha \beta } (C\gamma_{5} )_{\gamma \delta } \ ,
\label{nbardef}
\end{eqnarray}
where $C$ denotes the charge conjugation operator and $Q$ is a
Wuppertal-smeared \cite{wuppcite} quark field (with the superscript
denoting flavor), constructed iteratively as (where the superscript
now momentarily labels iterations):
\begin{equation}
Q^{(i)} (x) = (1-6\sigma ) Q^{(i-1)} (x) + \sigma \sum_{\mu =\pm 1}^{\pm 3}
U^{\dagger }_{\mu } (x-e_{\mu } ) Q^{(i-1)} (x-e_{\mu } ) \ .
\label{wupps}
\end{equation}
Here, $\sigma $ and the number of iterations $i_{max} $ are free parameters,
chosen such as to generate a good overlap between the neutron source and the
true neutron ground state \cite{wuppopt}. The iteration starts at
$Q^{(0)} \equiv \Psi $ and ends at $Q^{(i_{max} )} \equiv Q$. The sum
over directions $\mu $ in (\ref{wupps}) runs only over the three spatial
dimensions, but includes terms associated with both positive and negative
displacements in each dimension (i.e., $e_{-\mu } = -e_{\mu } $,
$U^{\dagger }_{-\mu } (x-e_{-\mu } ) = U_{\mu } (x)$). Note that smearing
constitutes a linear operation on the quark fields, i.e., there exists a
matrix $P$ such that
\begin{equation}
Q(x)=P(x,y) \Psi (y) \ ,
\ \ \ \ \ \ \ \ \ \ \ \ \ \ \ \ \
\bar{Q} (x)= \bar{\Psi } (y) P^{\dagger } (y,x) \ .
\label{smearmat}
\end{equation}
$P$ is proportional to the unit matrix in the Dirac indices, but not in
the space-time and color indices, nor in the flavor indices once the
external electric field is introduced via the substitution (\ref{amuintro}).
Note, thus, that the presence of the external electric field can influence
the smearing if one insists on manifest invariance of the neutron sources
and sinks with respect to gauge transformations of the external field.
However, it is not imperative to preserve such manifest invariance; not
doing so merely corresponds to evaluating (gauge-invariant) physical
observables in a particular gauge. In the treatment to follow, the most
general case will be considered, i.e., the perturbative expansion discussed
below will include the diagrams resulting from expanding the source
and sink fields in the external field. This will make it possible to
separately assess the influence of such terms. Ultimately, unambiguous
extraction of the neutron electric polarizability will be seen to
necessitate discarding such diagrams, and thus foregoing manifest
invariance of the neutron sources and sinks with respect to gauge
transformations of the external field; nevertheless, it will be verified
that the effect of including additional smearing diagram contributions
on the final result for the polarizability is negligible, thus rendering
this issue moot in any case.

\subsection{Perturbative expansion}
Having defined all of the objects entering the neutron two-point
function (\ref{2ptdef}), one can proceed to extract the quadratic term
of its Taylor expansion with respect to the external field $A_{\mu } $.
Both the interaction $S_E $ and the smeared neutron sources $N$ and
$\bar{N} $ in general contain a dependence on $A_{\mu } $. Expanding
(\ref{4dcoup}), one obtains two relevant vertices,
\begin{equation}
S_E = S_{E,1} + S_{E,2} + O\left( A_{\mu }^{3} \right) \ ,
\label{vertic}
\end{equation}
which can be written as bilinear forms,
\begin{equation}
S_{E,i} = \bar{\Psi } M_i \Psi \  ,
\label{bilin}
\end{equation}
with
\begin{eqnarray}
M_1 (x,y) &=& -iaz_V q_f \sum_{\mu } \left(
\frac{1-\gamma_{\mu } }{2} A_{\mu } (x) U_{\mu } (x)
\delta (x+e_{\mu } , y) \right. \label{se1} \\
& & \ \ \ \ \ \ \ \ \ \ \ \ \ \ \ \ \
-\left. \frac{1+\gamma_{\mu } }{2}
A_{\mu } (x-e_{\mu } ) U^{\dagger }_{\mu } (x-e_{\mu } )
\delta (x-e_{\mu } , y) \right) \nonumber \\
M_2 (x,y) &=& \frac{a^2 }{2} z_V q_f^2 \sum_{\mu } \left(
\frac{1-\gamma_{\mu } }{2} A_{\mu }^{2} (x) U_{\mu } (x)
\delta (x+e_{\mu } , y) \right. \label{se2} \\
& & \ \ \ \ \ \ \ \ \ \ \ \ \ \ \ \ \
+\left. \frac{1+\gamma_{\mu } }{2}
A_{\mu }^{2} (x-e_{\mu } ) U^{\dagger }_{\mu } (x-e_{\mu } )
\delta (x-e_{\mu } , y) \right) \ . \nonumber
\end{eqnarray}
Thus, $M_1 $ and $M_2 $ are matrices in the space-time, color, Dirac and
flavor indices, summation over which is implied in (\ref{bilin}).

On the other hand, also the smeared fields defined by (\ref{wupps})
need to be expanded in the external field,
\begin{equation}
Q^{(i)} = Q^{(i)}_{0} + Q^{(i)}_{1} + Q^{(i)}_{2}
+ O \left( A_{\mu }^{3} \right)
\end{equation}
(where the subscript denotes the order in the external field). Modifying
the link variables in (\ref{wupps}) according to (\ref{amuintro}) and
expanding in $A_{\mu } $, one has an iterative construction of the
smeared fields separated order by order in the external gauge field,
\begin{eqnarray}
Q^{(i)}_{0} (x) &=& (1-6\sigma ) Q^{(i-1)}_{0} (x)
+ \sigma \sum_{\mu =\pm 1}^{\pm 3} U^{\dagger }_{\mu } (x-e_{\mu } )
Q^{(i-1)}_{0} (x-e_{\mu } ) \label{ords0} \\
Q^{(i)}_{1} (x) &=& (1-6\sigma ) Q^{(i-1)}_{1} (x)
+ \sigma \sum_{\mu =\pm 1}^{\pm 3} U^{\dagger }_{\mu } (x-e_{\mu } )
\left( Q^{(i-1)}_{1} (x-e_{\mu } ) \right. \label{ords1} \\
& & \hspace{7cm} \left. -iaq_f A_{\mu } (x-e_{\mu } )
Q^{(i-1)}_{0} (x-e_{\mu } ) \right) \nonumber \\
Q^{(i)}_{2} (x) &=& (1-6\sigma ) Q^{(i-1)}_{2} (x)
+ \sigma \sum_{\mu =\pm 1}^{\pm 3} U^{\dagger }_{\mu } (x-e_{\mu } )
\left( Q^{(i-1)}_{2} (x-e_{\mu } ) \right. \label{ords2} \\
& & \hspace{7cm} -iaq_f A_{\mu } (x-e_{\mu } )
Q^{(i-1)}_{1} (x-e_{\mu } ) \nonumber \\ 
& & \hspace{7cm} \left. -\frac{a^2 q_f^2 }{2} A_{\mu }^{2} (x-e_{\mu } )
Q^{(i-1)}_{0} (x-e_{\mu } ) \right) \ . \nonumber
\end{eqnarray}
Equivalently, the smearing matrix $P$ in (\ref{smearmat}) can be
written in expanded fashion,
\begin{equation}
P = P_0 + P_1 + P_2 + O \left( A_{\mu }^{3} \right)
\label{pord}
\end{equation}
(the original quark field $\Psi $ is of course of zeroth order in 
$A_{\mu } $).

Returning to the neutron two-point function, expanding (\ref{2ptfrac})
in powers of $S_E $, inserting (\ref{vertic}) and discarding terms
which contribute only at higher than quadratic order in the external
field yields
\begin{eqnarray}
\langle N_{\alpha^{\prime } } (x^{\prime } ) \bar{N}_{\alpha } (x) \rangle
&=& \left\langle \left( 1-S_E + S_E^2 /2 \right)
N_{\alpha^{\prime } } (x^{\prime } ) \bar{N}_{\alpha } (x) \right\rangle_{0}
\left( 1+ \langle S_E - S_E^2 /2 \rangle_{0}
+ \langle S_E \rangle_{0}^{2} \right) \nonumber \\
&=& \langle N_{\alpha^{\prime } } (x^{\prime } )
\bar{N}_{\alpha } (x) \rangle_{0} \nonumber \\
& & \ \ \ \ \ \ \ \ -
\left\langle \left( S_{E,1} + S_{E,2} - S_{E,1}^{2} /2 \right)
N_{\alpha^{\prime } } (x^{\prime } ) \bar{N}_{\alpha } (x)
\right\rangle_{0} \nonumber \\
& & \ \ \ \ \ \ \ \ +
\left\langle S_{E,1} + S_{E,2} - S_{E,1}^{2} /2 \right\rangle_{0}
\left\langle N_{\alpha^{\prime } } (x^{\prime } ) \bar{N}_{\alpha } (x)
\right\rangle_{0} \label{2ptexp} \\
& & \ \ \ \ \ \ \ \ -
\left\langle S_{E,1} \right\rangle_{0}
\left\langle S_{E,1} N_{\alpha^{\prime } } (x^{\prime } )
\bar{N}_{\alpha } (x) \right\rangle_{0} \nonumber \\
& & \ \ \ \ \ \ \ \ +
\left\langle S_{E,1} \right\rangle_{0}^{2} \left\langle
N_{\alpha^{\prime } } (x^{\prime } ) \bar{N}_{\alpha } (x)
\right\rangle_{0} \ . \nonumber
\end{eqnarray}
As usual, the denominator in the original expression (\ref{2ptfrac}) has
the effect of subtracting disconnected (in the statistical sense) pieces.
Furthermore, inserting the more specific forms (\ref{ndef}), (\ref{nbardef}),
(\ref{smearmat}) and (\ref{bilin}), one arrives at (the superscripts of the
smearing matrices $P$ and the quark fields $\Psi $ denoting a fixed flavor):
\begin{eqnarray}
\langle N_{\alpha^{\prime } } (x^{\prime } ) \bar{N}_{\alpha } (x) \rangle
\! \! &=& \! \! \delta_{\alpha^{\prime } \beta^{\prime } }
(C\gamma_{5} )_{\gamma^{\prime } \delta^{\prime } }
\epsilon_{b^{\prime } c^{\prime } d^{\prime } }
\epsilon_{bcd} \delta_{\alpha \beta } (C\gamma_{5} )_{\gamma \delta }
\nonumber \\
& & \! \! \times
P^{(d)}_{b^{\prime } k^{\prime } } (x^{\prime },u^{\prime })
P^{(d)}_{c^{\prime } l^{\prime } } (x^{\prime },v^{\prime })
P^{(u)}_{d^{\prime } m^{\prime } } (x^{\prime },w^{\prime })
P^{\dagger (u)}_{md} (w,x) P^{\dagger (d)}_{lc} (v,x)
P^{\dagger (d)}_{kb} (u,x) \nonumber \\
& & \! \! \times \left[ -\left\langle
\Psi^{(d)}_{k^{\prime } \beta^{\prime } } (u^{\prime } )
\Psi^{(d)}_{l^{\prime } \gamma^{\prime } } (v^{\prime } )
\Psi^{(u)}_{m^{\prime } \delta^{\prime } } (w^{\prime } )
\bar{\Psi }^{(u)}_{m\delta } (w) \bar{\Psi }^{(d)}_{l\gamma } (v)
\bar{\Psi }^{(d)}_{k\beta } (u) \right. \right.
\label{2ptlong} \\
& & \ \ \ \ \ \ \ \ \ \ \ \ \ \ \ \ \ \ \ \ \ \ \ \times \left.
\left( (\bar{\Psi } M_1 \Psi ) + (\bar{\Psi } M_2 \Psi )
-\frac{1}{2} (\bar{\Psi } M_1 \Psi ) (\bar{\Psi } M_1 \Psi ) \right)
\right\rangle_{0} \nonumber \\
& & \ \ \ -\left\langle
\Psi^{(d)}_{k^{\prime } \beta^{\prime } } (u^{\prime } )
\Psi^{(d)}_{l^{\prime } \gamma^{\prime } } (v^{\prime } )
\Psi^{(u)}_{m^{\prime } \delta^{\prime } } (w^{\prime } )
\bar{\Psi }^{(u)}_{m\delta } (w) \bar{\Psi }^{(d)}_{l\gamma } (v)
\bar{\Psi }^{(d)}_{k\beta } (u) \ (\bar{\Psi } M_1 \Psi ) \right\rangle_{0}
\nonumber \\
& & \hspace{9.3cm} \times \left\langle
(\bar{\Psi } M_1 \Psi ) \right\rangle_{0} \nonumber \\
& & \ \ \ +\left\langle
\Psi^{(d)}_{k^{\prime } \beta^{\prime } } (u^{\prime } )
\Psi^{(d)}_{l^{\prime } \gamma^{\prime } } (v^{\prime } )
\Psi^{(u)}_{m^{\prime } \delta^{\prime } } (w^{\prime } )
\bar{\Psi }^{(u)}_{m\delta } (w) \bar{\Psi }^{(d)}_{l\gamma } (v)
\bar{\Psi }^{(d)}_{k\beta } (u) \right\rangle_{0} \nonumber \\
& & \ \ \ \ \ \ \ \ \ \ \ \ \ \ \ \times \left( 1+
\left\langle (\bar{\Psi } M_1 \Psi ) + (\bar{\Psi } M_2 \Psi )
-\frac{1}{2} (\bar{\Psi } M_1 \Psi ) (\bar{\Psi } M_1 \Psi )
\right\rangle_{0} \right. \nonumber \\
& & \hspace{8.9cm}
+ \left. \left. \left\langle (\bar{\Psi } M_1 \Psi )
\right\rangle_{0}^{2} \right) \right] \nonumber
\end{eqnarray}
Applying Wick's theorem (i.e., evaluating the $\langle \ldots \rangle_{U} $
averages over the quark fields, cf.~(\ref{milcav}),(\ref{ou})), and
retaining only contributions quadratic in the external gauge field, one
finally arrives at a diagrammatic representation, depicted in
Fig.~\ref{fig4pt}, for the desired quantity, namely, the quadratic
term in the Taylor expansion of the neutron two-point function with
respect to the external field. The diagrams in Fig.~\ref{fig4pt} are to
be read as follows:

\begin{figure}
\centerline{\epsfig{file=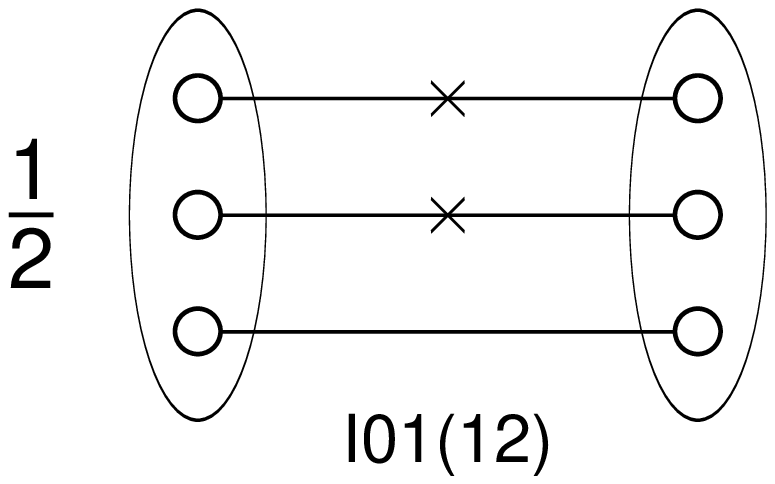,width=3.9cm}
\epsfig{file=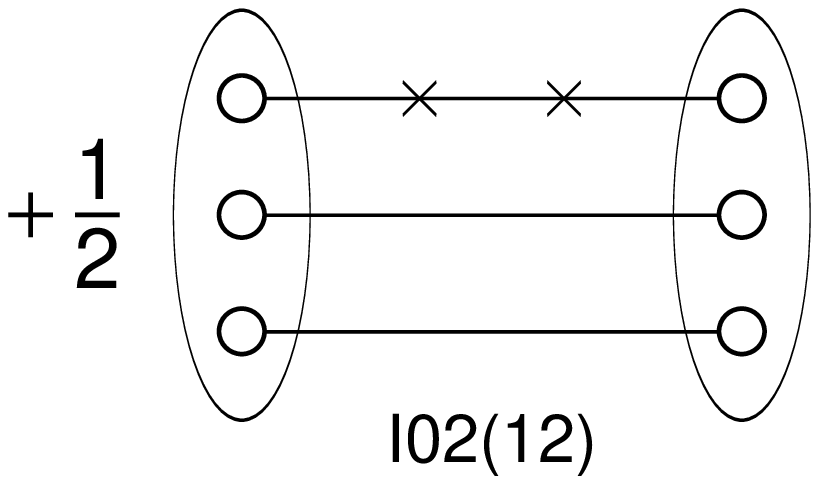,width=3.9cm}
\epsfig{file=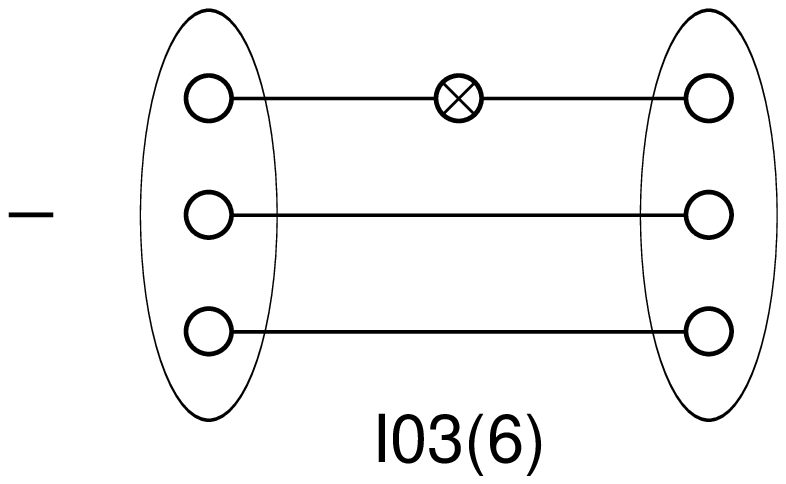,width=3.9cm}
\epsfig{file=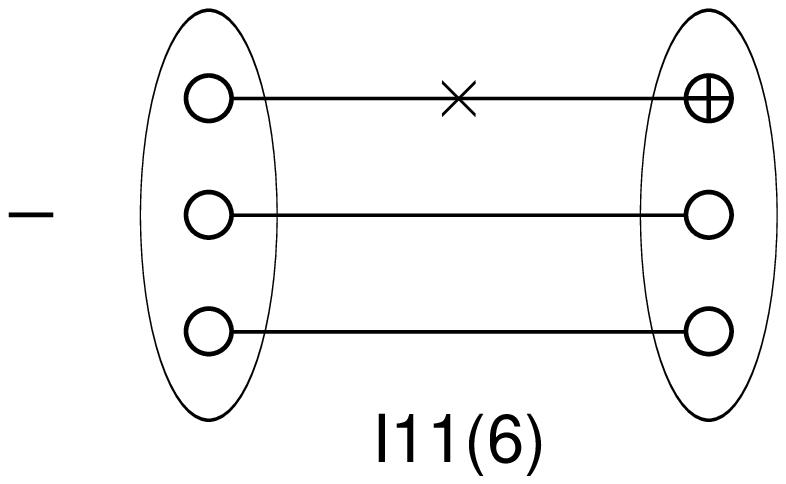,width=3.9cm} }
\centerline{\epsfig{file=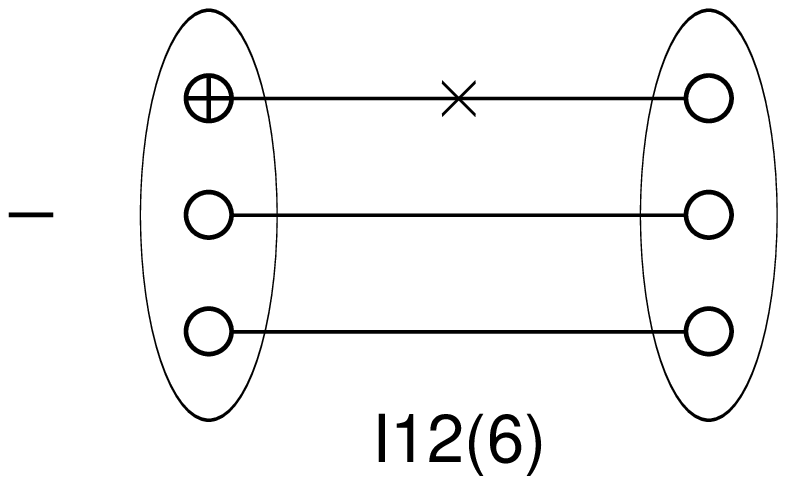,width=3.9cm}
\epsfig{file=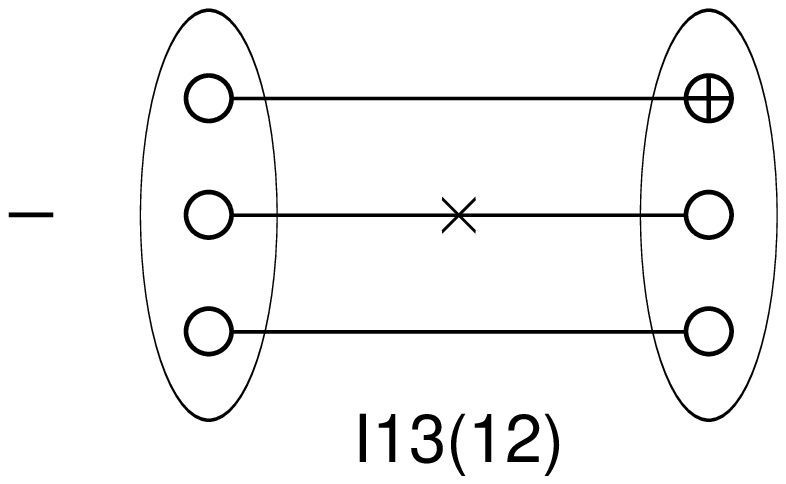,width=3.9cm}
\epsfig{file=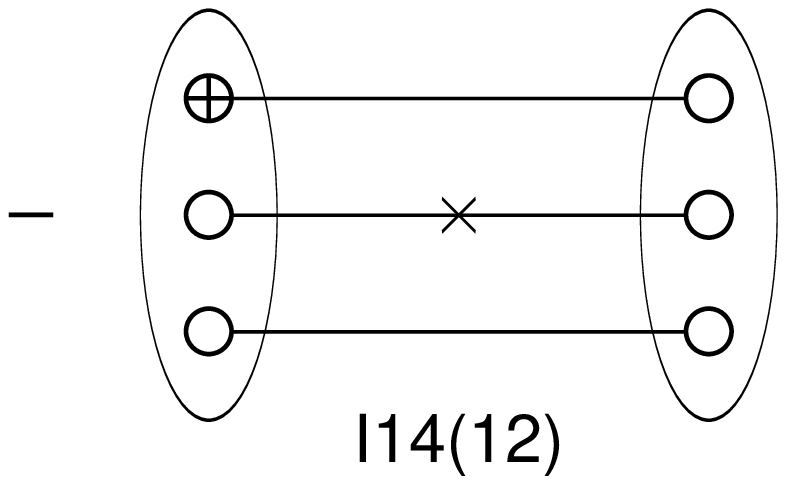,width=3.9cm}
\epsfig{file=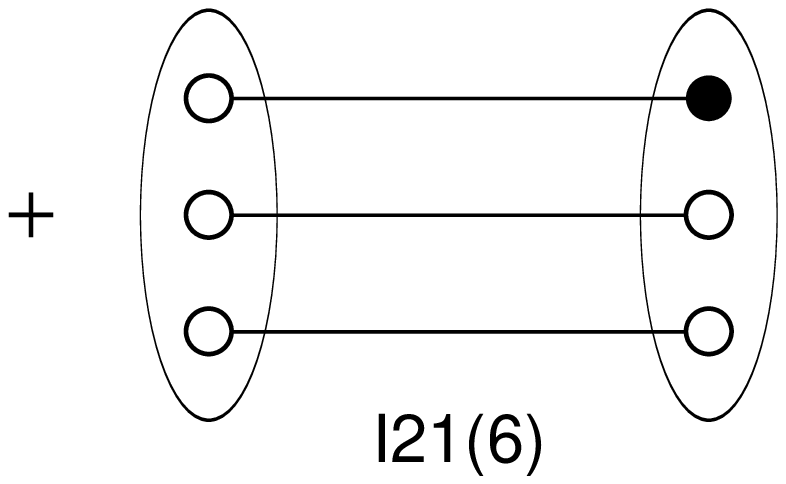,width=3.9cm} }
\centerline{\epsfig{file=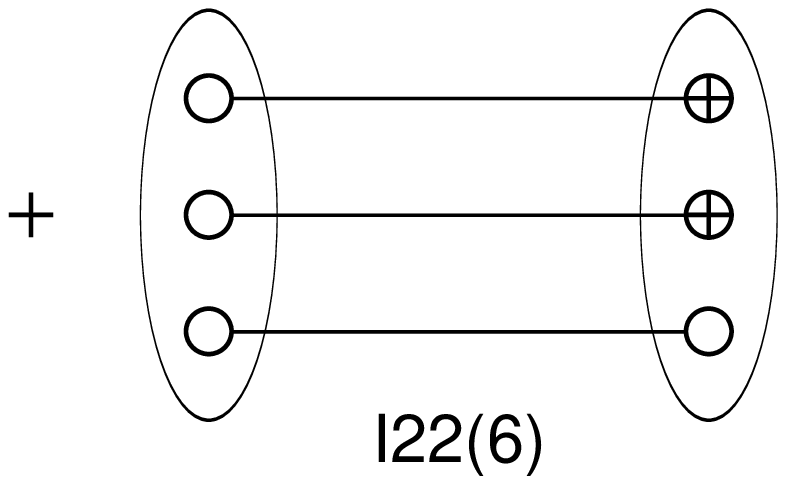,width=3.9cm}
\epsfig{file=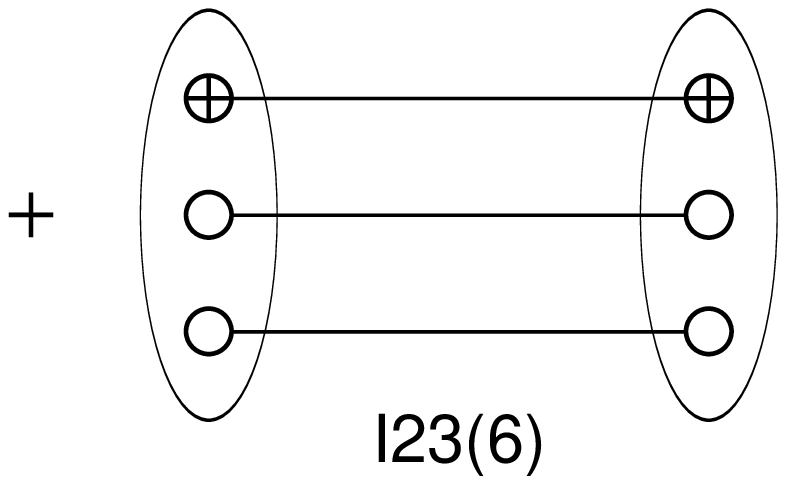,width=3.9cm}
\epsfig{file=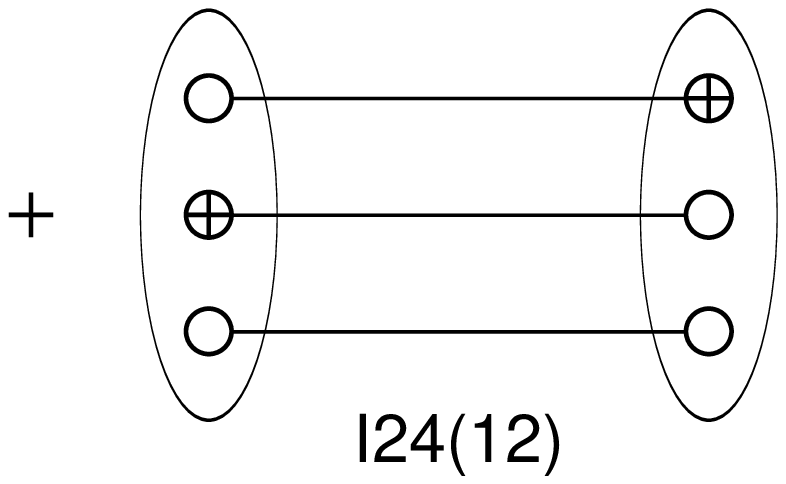,width=3.9cm}
\epsfig{file=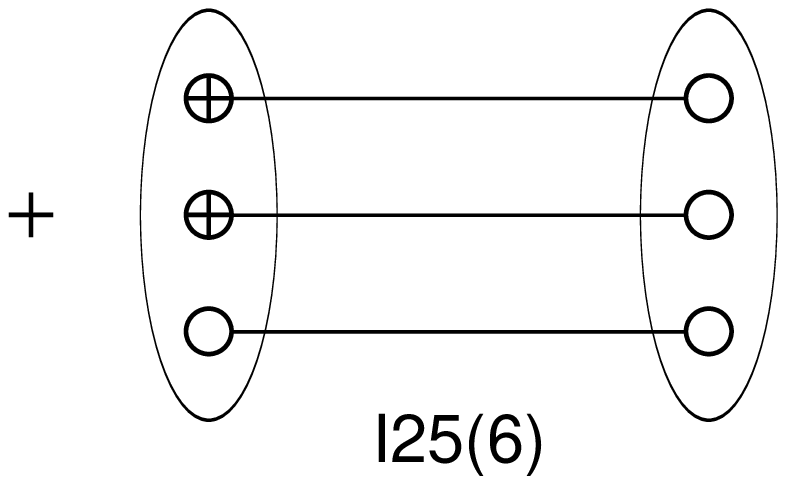,width=3.9cm} }
\vspace{1.3cm}
\centerline{\epsfig{file=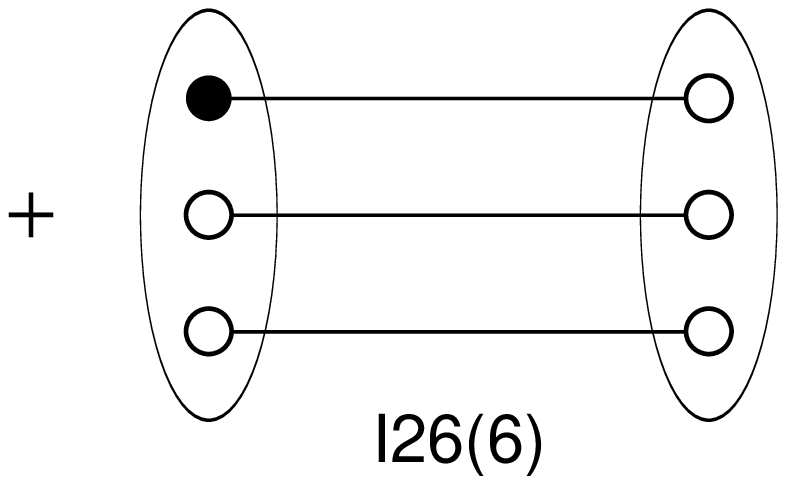,width=3.9cm}
\epsfig{file=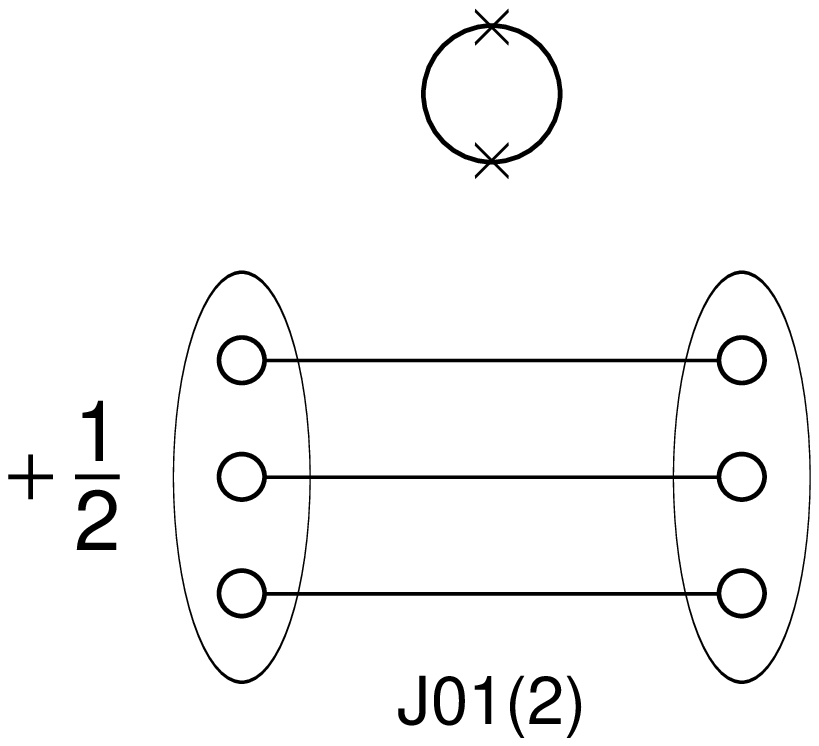,width=3.9cm}
\epsfig{file=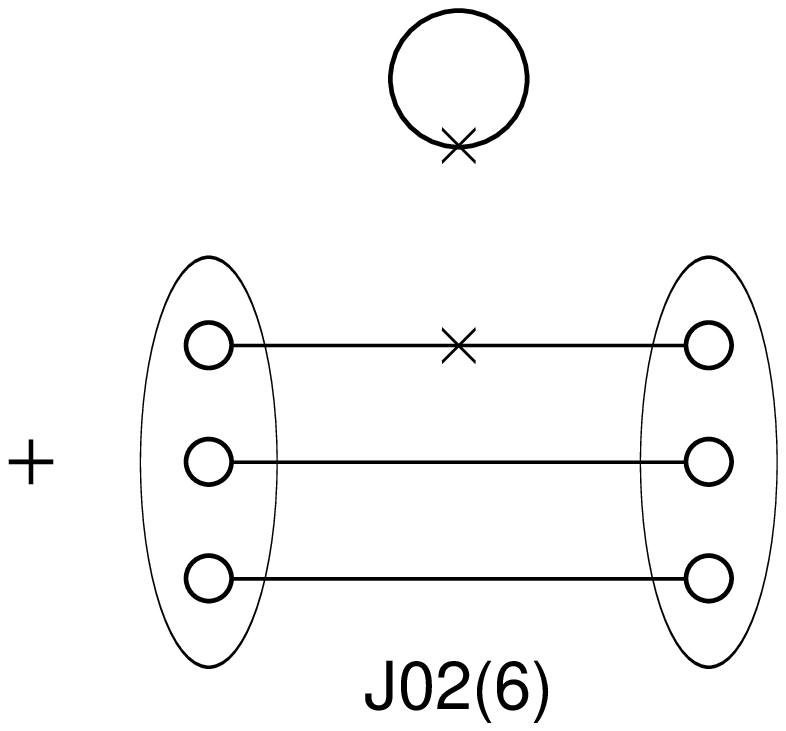,width=3.9cm}
\epsfig{file=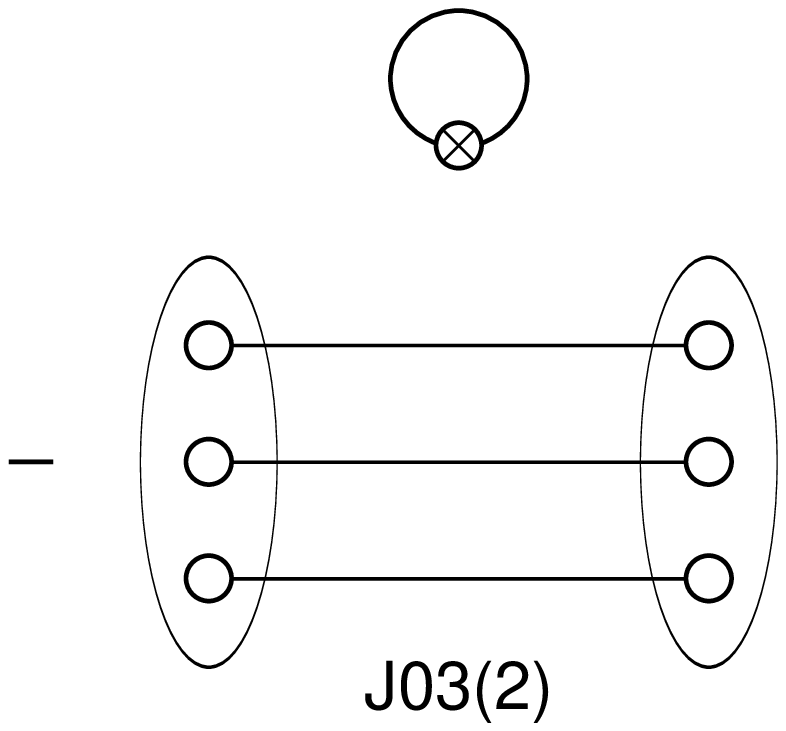,width=3.9cm} }
\vspace{1.3cm}
\centerline{\epsfig{file=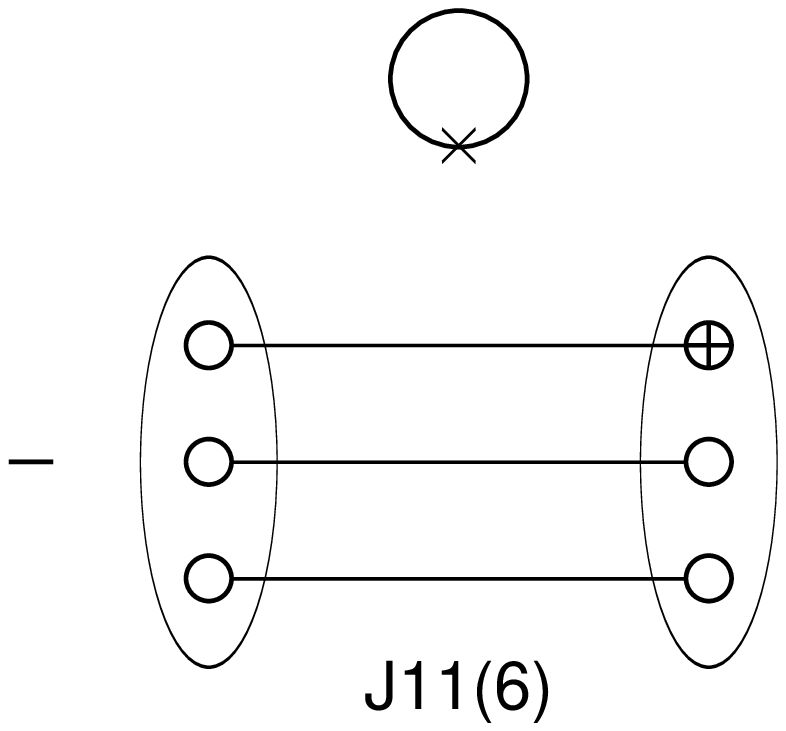,width=3.9cm}
\epsfig{file=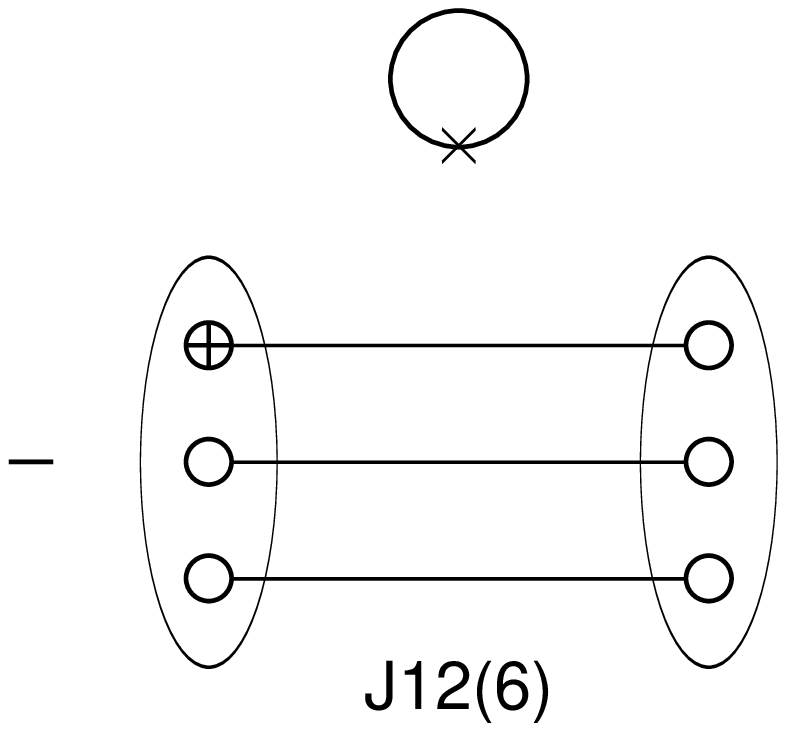,width=3.9cm}
\epsfig{file=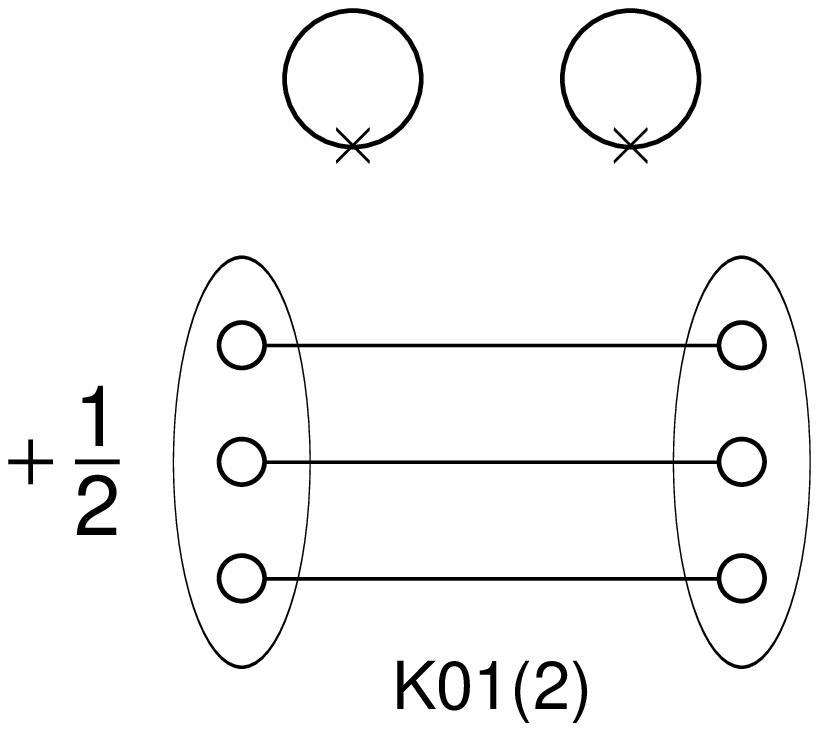,width=3.9cm} }
\caption{Contributions to the neutron two-point function quadratic in
the external gauge field. The nomenclature is explained in detail
in the main text.}
\label{fig4pt}
\end{figure}

\begin{enumerate}
\item[a.] Solid lines are point-to-point quark propagators
$K^{(f)\, c^\prime c}_{\gamma^{\prime } \gamma } (x^{\prime } ,x) =
\langle \Psi^{(f)}_{c^{\prime } \gamma^{\prime } } (x^{\prime } )
\bar{\Psi}^{(f)}_{c\gamma } (x) \rangle_{U} $. Note that these are
propagators between four-dimensional sources and sinks, i.e., an
initially four-dimensional source is propagated in five dimensions
using the domain wall quark action (\ref{sdwf}) and finally projected
back to four-dimensional space-time according to the correspondence
(\ref{projq4}). Of the three propagator chains connecting neutron
source and sink, two are associated with flavor down, $f=d$, and one
with flavor up, $f=u$. The quark loops imply a sum over all three
flavors.
\item[b.] \parbox{1cm}{\Huge $\circ $} \hspace{-0.7cm} denotes
quark source or sink smearing of zeroth order in the external field,
i.e. $P_0 $ in the decomposition (\ref{pord}). Similarly,
\parbox{1cm}{$\bigoplus $ \vspace{0.08cm} } \hspace{-0.74cm} corresponds
to $P_1 $ and \parbox{1cm}{\Huge $\bullet $} \hspace{-0.74cm} corresponds
to $P_2 $.
\item[c.] {\Large $\times $} denotes a vertex insertion linear in the
external field, i.e., multiplication by $M_1 $, cf.~(\ref{se1}). Similarly,
\parbox{1cm}{$\bigotimes $ \vspace{0.07cm} } \hspace{-0.74cm} corresponds
to $M_2 $, cf.~(\ref{se2}).
\item[d.] At the neutron source and sink, symbolized by the ovals, color
and Dirac indices must be contracted in accordance with the first line
of the right-hand side of (\ref{2ptlong}).
\item[e.] Each diagram summarizes several elementary terms in the
Wick expansion of (\ref{2ptlong}). For every contribution in which the
quark lines run literally as shown, there is a corresponding contribution
in which the two down quark lines connecting neutron source and sink
cross (i.e., the sinks are exchanged). The latter contribution receives
an additional minus sign from the exchange. Furthermore, for each diagram,
there are several ways of distributing vertices and smearings over the
quark sources, sinks and propagators, only one of which is shown in each
case. Note that the combinatorics are different for vertices and
smearings. On the one hand, there are six ways of distributing two
{\Large $\times $} vertices such that they reside on different quark
lines connecting neutron source and sink (i.e., exchanging two such vertices
amounts to a new contribution\footnote{To be completely precise, this only
applies when both vertices reside on quark lines connecting neutron source
and sink; on the other hand, in the diagrams labeled $J01(2)$, $J02(6)$
and $K01(2)$, no additional contributions stemming from exchange of the
vertices are implied. Any such duplications which may arise from the
Wick expansion of (\ref{2ptlong}) are already taken into account through
the prefactor of the diagram (such a duplication actually only occurs in
the case of $J02(6)$, the statistically connected part of which,
cf.~item~g., originally enters with a prefactor of $1/2$).});
on the other hand, there are only three ways of distributing two
\parbox{1cm}{$\bigoplus $ \vspace{0.08cm} } \hspace{-0.74cm}
sink smearings in the neutron sink (since the product
$(P_0 +P_1 +P_2 )^3 $ contains only three terms consisting of two
factors $P_1 $ and one factor $P_0 $), and analogously for the neutron
source. Note that the labeling of the diagrams
reflects these multiplicities; in each label, the integer inside the
parentheses denotes the number of individual contributions from the
Wick expansion of (\ref{2ptlong}) summarized by the diagram.
\item[f.] As usual, each quark loop implies an additional minus sign.
In order to keep with standard nomenclature, these signs were not
absorbed into the prefactors, but must be included separately when
evaluating the diagrams.
\item[g.] Each contribution finally must be averaged over the gauge
ensemble, where, as already remarked after eq.~(\ref{2ptexp}), statistically
disconnected parts are subtracted. Thus, denoting the gauge ensemble
average as
\begin{equation}
\langle O \rangle_{G} = \frac{1}{Z_{0} } \int [DU]
\exp (-S_{G,eff} ) O \ ,
\end{equation}
diagram $J01(2)$ is to be evaluated as
\begin{figure}[h]
\centerline{\epsfig{file=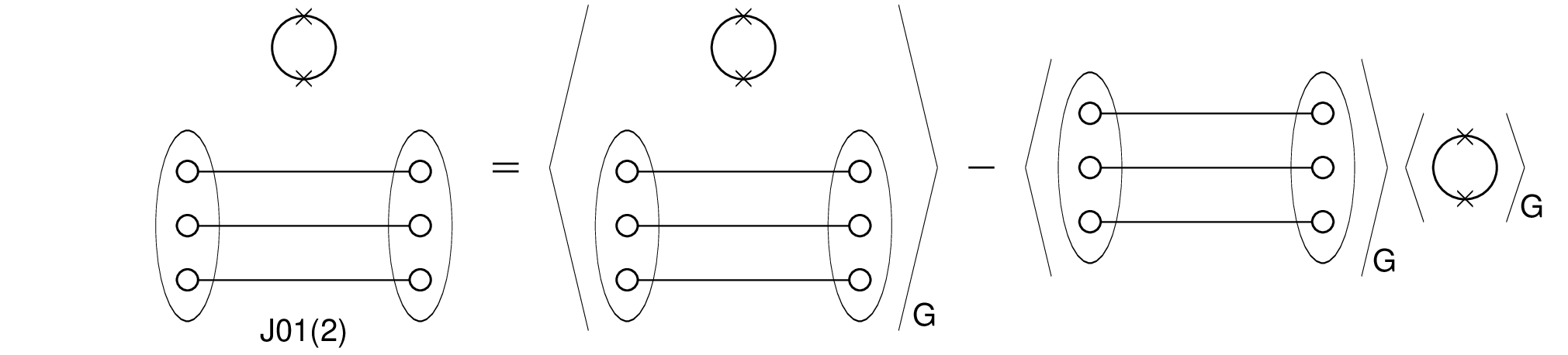,width=14cm} }
\end{figure}

(where of course only items a.-f.~apply to the objects inside the averages).
Diagrams $J02(6)$, $J03(2)$, $J11(6)$ and $J12(6)$ are treated analogously.
The more complicated case $K01(2)$ is evaluated as
\begin{figure}[h]
\centerline{\epsfig{file=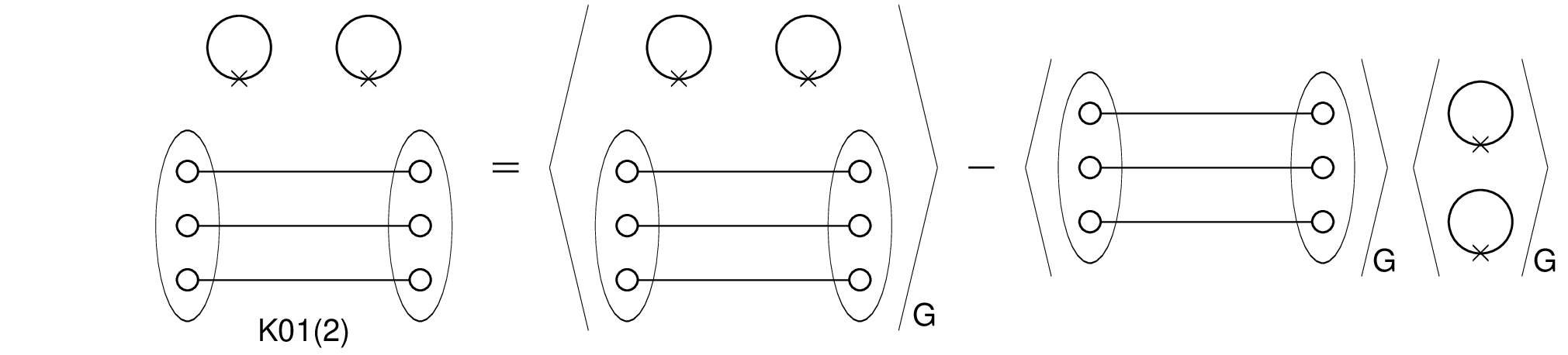,width=14cm} 
\hspace{2.3cm} \vspace{0.5cm} }
\centerline{\hspace{2cm} \epsfig{file=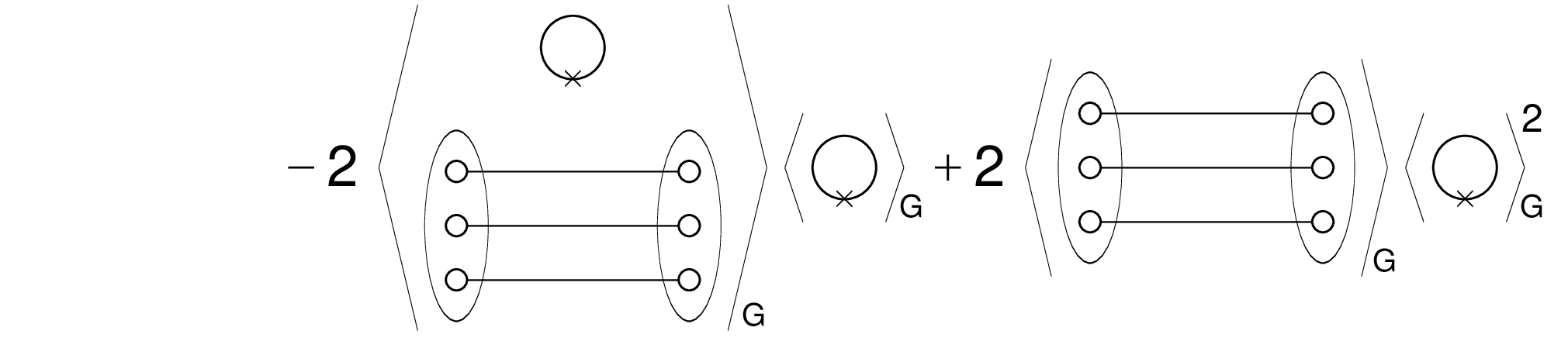,width=14cm} }
\vspace{-0.5cm}
\end{figure}
\end{enumerate}

Note that the naming of the different contributions is intended to be
mnemonic. The initial character differentiates between contributions
of varying number of disconnected parts; $I$ denotes connected diagrams,
$J$ disconnected ones with two parts and $K$ the disconnected diagram
with three parts. The digit following the initial character indicates the
power of the external electric field contributed specifically by the source
and sink smearings. The next digit is simply a running index numbering the
contributions in each class. Finally, as already mentioned under
item e.~above, the integer in the parentheses denotes the number of
individual contributions from the Wick expansion of (\ref{2ptlong})
summarized by the diagram.

\subsection{Calculational details}
\label{calcsec}
The code written to compute the diverse diagrams in Fig.~\ref{fig4pt}
relied heavily on the Chroma Library for Lattice Field Theory
\cite{chrom1,chrom2}. In practice, the propagator chains connecting neutron
source and sink were calculated in sequential fashion. Starting from
a space-time location $x$ and a specific set of color, Dirac and flavor
indices $a$, $\alpha $ and $f$, as well as choosing the desired order $i$
of the smearing in the external field, one constructs the smeared source
vector $\delta_{\beta \alpha } P_{i\ \ ba}^{\dagger (f)} (y,x)$ using the
appropriate iterative procedure (\ref{ords0}), (\ref{ords1}) or
(\ref{ords2}). While no loss of generality is incurred by
performing the calculation for only one particular $x$, all
combinations of the other indices are ultimately required for the
contractions at the neutron source (of course, different flavors are
related in a trivial manner). Propagating the aforementioned specific
smeared source vector yields directly the smeared-to-point propagator
$K^{(f)\, cb}_{\gamma \alpha } (z,y) P_{i\ \ ba}^{\dagger (f)} (y,x)$.
A vertex insertion implies multiplication with the corresponding matrix
$M_j $, yielding a new source vector $M_{j\ \ \delta \gamma }^{(f)\, dc}
(w,z) K^{(f)\, cb}_{\gamma \alpha } (z,y) P_{i\ \ ba}^{\dagger (f)} (y,x)$.
This source vector is then again propagated\footnote{Note that the
positions of the interaction vertices in the diagrams in Fig.~\ref{fig4pt}
are not external parameters, but integration variables. Thus, e.g.,
diagram $I01(12)$ does not represent a full four-point function, but
only a very specific space-time integral over a four-point function.
It is these integrations which render the calculation tractable by the
sequential procedure described here; they provide precisely the
contraction between a vertex and an attached propagator which permits
treating an inserted vertex simply as one single new source, devoid of
external parameters and spread out over all of space-time, to be submitted
to the subsequent propagation.}, thus building up the propagator chain
sequentially. When finally arriving at the neutron sink, the appropriate
sink smearing is applied, using again (\ref{ords0}), (\ref{ords1}) or
(\ref{ords2}).

The disconnected quark loops were evaluated using stochastic estimation.
To estimate the trace over all indices implied by the loop, a basis of
120 stochastic sources (240 for two cases of external fields which
engender particularly strong statistical fluctuations, cf.~section
\ref{massres}) was used. Again, starting at each stochastic source,
propagator chains were constructed sequentially, and finally contracted
again with the stochastic source. The sources were complex $Z(2)$ sources,
distributed homogeneously over space-time, Dirac and color space\footnote{For
the flavor $SU(3)$-symmetric ensemble $m_u =m_d =m_s $ used in this work, it
is sufficient to consider one flavor and weight the result by the appropriate
combination of fractional charges to obtain the full value of the loop
diagram.}, i.e., each point in that product space was associated with
a value from the set
\begin{equation}
\left\{ 1+i, 1-i, -1+i, -1-i \right\}
\end{equation}
with equal probability.

\begin{figure}[t]
\centerline{\hspace{0.2cm}
\epsfig{file=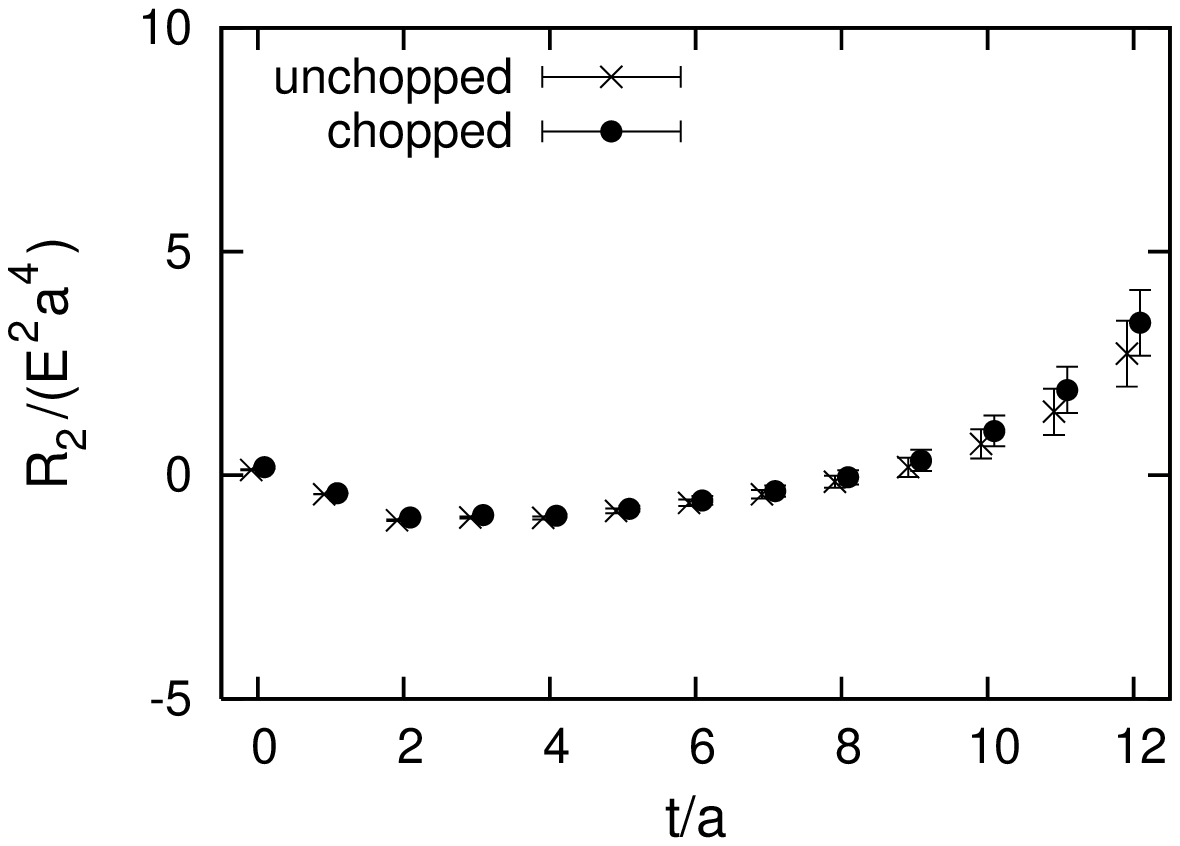,width=5.3cm,angle=-90}
\hspace{0.5cm}
\epsfig{file=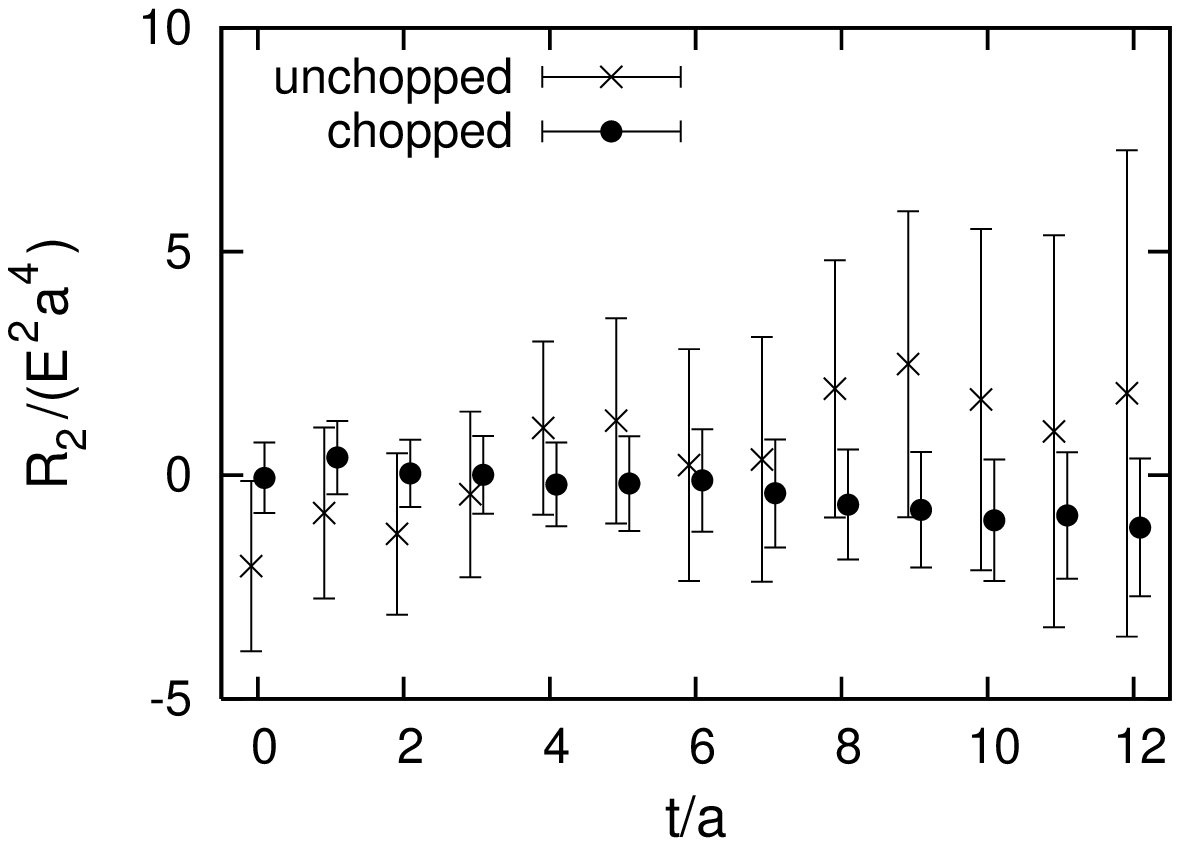,width=5.3cm,angle=-90} }
\caption{Comparison of results obtained using chopped and unchopped
external fields, as described in the main text. Left panel
displays the sum of the connected diagrams $I01$, $I02$ and $I03$;
right panel the sum of the disconnected diagrams $J01$ and $J03$. Results
are shown as a function of temporal source--sink separation, in each case
normalized by the neutron two-point function in the absence of the
external field, i.e., shown are the contributions by the respective
subsets of diagrams to the ratio $R_2 $ defined in eq.~(\ref{ratio2}). All
measurements are taken at integer times; data are slightly displaced from
those times in the figures for better readability. The electric field $E$
providing the scale is cast in Gau\ss ian units. Shown are unrenormalized
raw data, i.e., for the purpose of this comparison, $z_V =1$ in the
vertices (\ref{se1}),(\ref{se2}).}
\label{chopping1}
\end{figure}

\begin{figure}[t]
\centerline{\epsfig{file=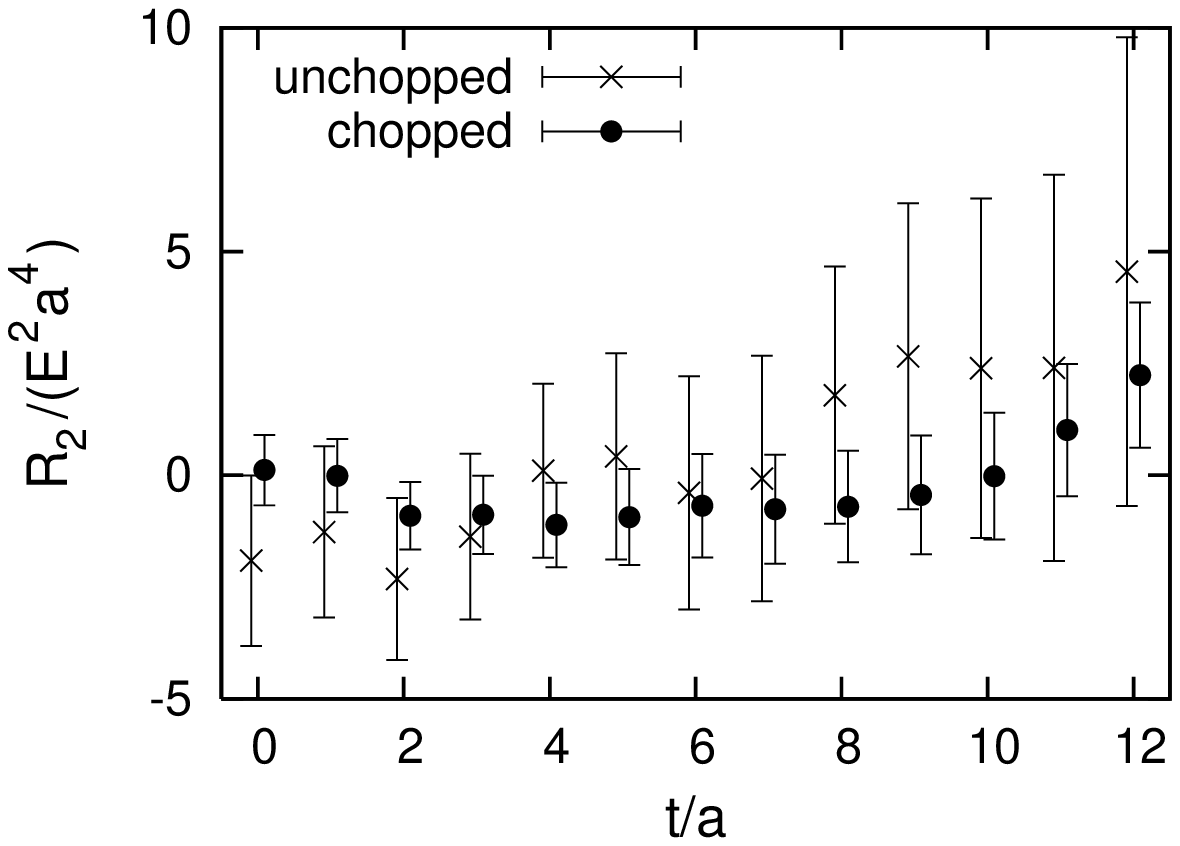,width=5.3cm,angle=-90} }
\caption{Comparison of results obtained using chopped and unchopped
external fields, analogous to Fig.~\ref{chopping1}, but showing
the contributions of the connected diagrams $I01$, $I02$, $I03$ and
the disconnected diagrams $J01$, $J03$ in one figure.}
\label{chopping2}
\end{figure}

Disconnected contributions exhibit strong statistical fluctuations, and
two possibilities of reducing these fluctuations were investigated. On the
one hand, the consequences of only switching on the external electric
field a short time before the introduction of the neutron source and
switching it off soon after the annihilation by the neutron sink were
explored. This procedure will be referred to as ``chopping'' the external
field in the following. It is motivated by the expectation that, if
sufficient time has elapsed between the introduction of the neutron source
and the neutron mass measurement to filter out the true neutron ground
state, then also any switching-on effects generated prior to the
introduction of the neutron source will have decayed. However, the
statistical fluctuations of disconnected diagrams will be significantly
affected by chopping the external field. Summing up contributions due
to the coupling of the external field to vacuum fluctuations far in the
past or the future of the neutron mass measurement, while not expected
to influence the outcome of the latter, will certainly add statistical
noise to it. Chopping the external field can reduce that noise
significantly by discarding irrelevant vacuum fluctuations. This was
tested using the external field $A_3 =E(t-t_0 )$ with $t_0 =-10a$,
where here and in the following, the temporal lattice boundaries are
located at $t=-10a$ and $t=22a$, and the neutron source is located at
$t=0$. Figs.~\ref{chopping1} and \ref{chopping2} compare results obtained
without chopping, i.e., $A_3 =E(t-t_0 )$ throughout the lattice, with
results obtained by setting $A_3 =E(t-t_0 )$ only for $-a\leq t\leq 14a$
and $A_3 =0$ for other times. As expected, no significant differences
arise in the measured correlator ratio. This is particularly clear in
the connected contributions, which are determined very accurately; in
the disconnected contributions, a significant reduction of the statistical
uncertainty results. Note that the neutron mass shift is ultimately
extracted specifically from the slope of the correlator ratio shown
in Figs.~\ref{chopping1} and \ref{chopping2}, as discussed in section
\ref{interpsec}. Due to the advantages offered by chopping the external
field, all further measurements reported in the following, cf.~in particular
section \ref{massres}, were obtained using chopped external electric fields.

\begin{figure}[t]
\centerline{\hspace{0.2cm}
\epsfig{file=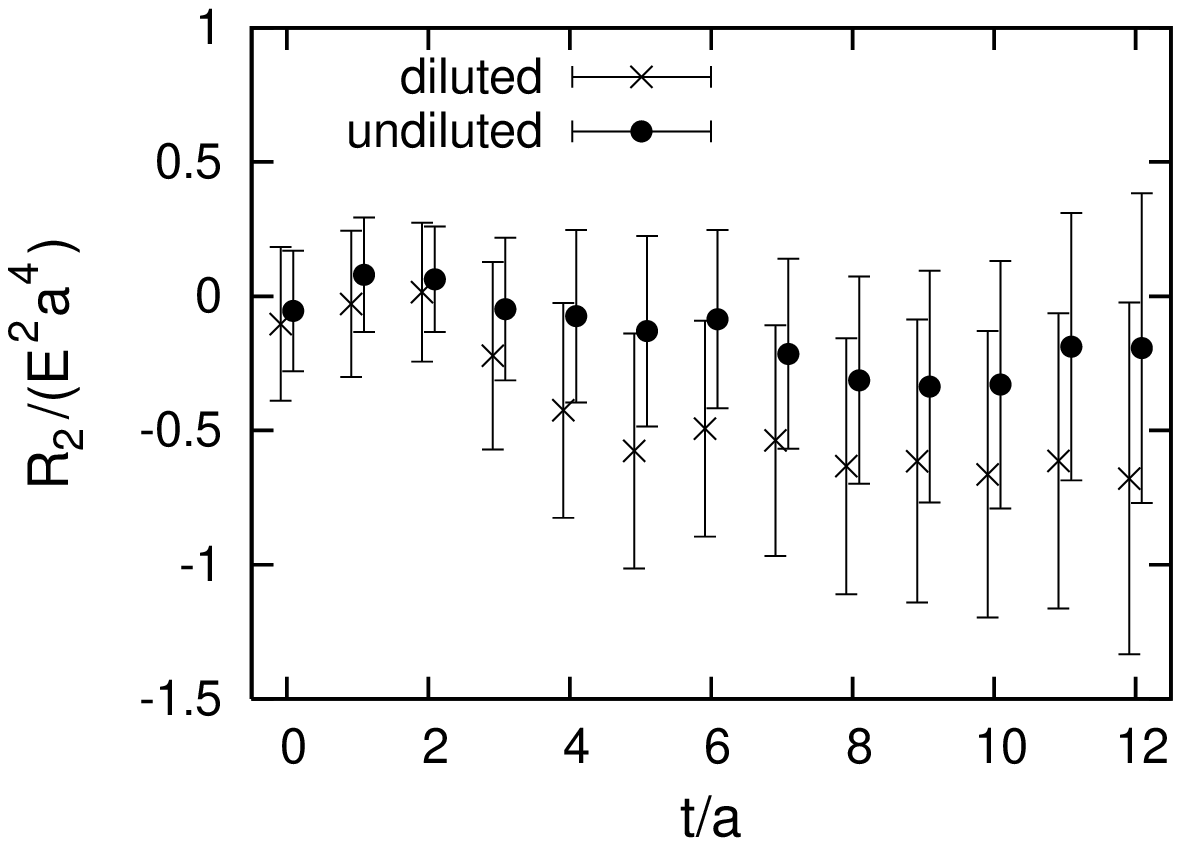,width=5.3cm,angle=-90}
\hspace{0.5cm}
\epsfig{file=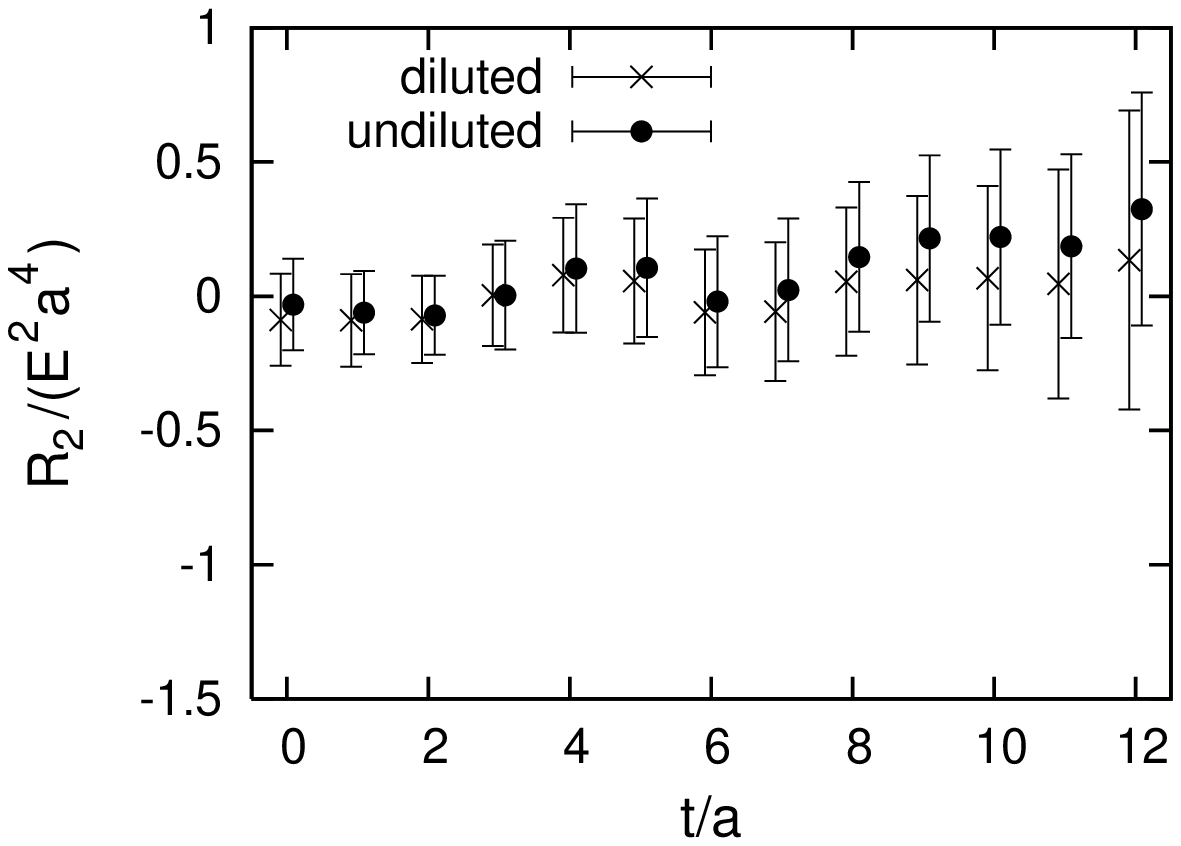,width=5.3cm,angle=-90} }
\caption{Comparison of results obtained using undiluted and diluted
stochastic sources, as described in the main text. Diagram $J01$ (left)
and diagram $J03$ (right) are displayed as a function of temporal
source--sink separation, each normalized by the neutron two-point function
in the absence of the external field, i.e., shown are the individual
contributions by the two diagrams to the ratio $R_2 $ defined in
eq.~(\ref{ratio2}). All measurements are taken at integer times;
data are slightly displaced from those times in the figures for
better readability. The electric field $E$ providing the scale is
cast in Gau\ss ian units. Shown are unrenormalized raw data, i.e.,
for the purpose of this comparison, $z_V =1$ in the vertices
(\ref{se1}),(\ref{se2}).}
\label{dilution}
\end{figure}

On the other hand, a further possibility of reducing the uncertainty of
stochastic estimation which was explored is dilution \cite{dublin},
specifically dilution in the Dirac index. In other words, besides the
stochastic estimation scheme described above, also an alternative scheme
was considered in which each value of the Dirac index in the loop trace
was considered separately, with $Z(2)$ sources distributed homogeneously
only over space-time and color space in each case, the sum over those values
yielding the Dirac trace at the end. The comparison between the two schemes
was carried out for the external field $A_3 =Et$, chopped as described
further above, with the neutron source again located at $t=0$.
Fig.~\ref{dilution} shows the respective results obtained for the
disconnected diagrams $J01$ and $J03$. Evidently, for this particular
external field and dilution scheme, there is no computational advantage
in dilution; the statistical uncertainty in fact is slightly larger in
the diluted case. As a consequence, dilution was not considered any further
in the present investigation; this does not exclude that a comprehensive
survey of various implementations of dilution could yield computationally
more advantageous schemes.

\section{Interpretation of the neutron two-point function}
\label{interpsec}
The standard method of extracting ground-state hadron masses is to
project the hadron two-point function onto a definite momentum, consider
an appropriate Dirac component, and compare the measured data to the
corresponding spectral representation. Choosing, specifically, zero
momentum and unpolarized neutron states,
\begin{equation}
G(p=0,t) =
\int d^3 x^{\prime } \, \mbox{Tr} \, \left( \frac{1+\gamma_{0} }{2}
\langle N (x^{\prime } ) \bar{N} (x) \rangle \right) \longrightarrow
W\exp (-mt)
\label{2ptdecay}
\end{equation}
for sufficiently large times $t$, where the neutron source location
defines $t=0$, and $W$ characterizes the overlap between the state
created by the operator $\bar{N} $ and the true neutron ground state.
Thus, the neutron mass $m$ can be extracted from the exponential decay
of the correlator (\ref{2ptdecay}).

Furthermore, if one is calculating
the correlator as a function of a small external parameter, such as an
external electric field $E$, one can expand in $E$,
\begin{eqnarray}
m &=& m_0 + m_1 E + m_2 E^2 + O \left( E^3 \right) \\
W &=& W_0 + W_1 E + W_2 E^2 + O \left( E^3 \right)
\end{eqnarray}
and then the Taylor expansion of (\ref{2ptdecay}) contains the quadratic
term
\begin{equation}
G^{(2)} (p=0,t) = \exp (-m_0 t) \left[ W_2 -W_0 m_2 t -W_1 m_1 t
+W_0 m_1^2 t^2 /2 \right] E^2 \ .
\end{equation}
Using the fact that the neutron's electric dipole moment vanishes,
$m_1 =0$, and dividing by the correlator $G_0 $ obtained in the absence of
the external field,
\begin{equation}
G_0 (p=0,t) \longrightarrow W_0 \exp (-m_0 t) \ ,
\end{equation}
one has
\begin{equation}
R_2 (t) \equiv \frac{G^{(2)} (p=0,t)}{G_0 (p=0,t)} \longrightarrow
\left( \frac{W_2 }{W_0 } - m_2 t \right) E^2 \ ,
\label{ratio2}
\end{equation}
allowing one to extract the neutron electric polarizability
\begin{equation}
\alpha = -2m_2 \ ,
\end{equation}
cf.~(\ref{leadpol}), from the slope of (\ref{ratio2}) as a function of $t$.

Two assumptions underlie this procedure, namely, time-independence of
the Hamiltonian and spatial translational invariance. As already indicated
in section \ref{intsec}, neither of the external gauge fields (\ref{a3def})
and (\ref{a0def}) investigated in the present work satisfies both of these
assumptions simultaneously. As a result, the standard analysis discussed
above, which would be appropriate in infinitely extended space-time, needs
to be reconsidered in more detail.

\subsection{Temporally varying gauge field}
\label{a3sec}
Consider first the case of the external field (\ref{a3def}),
\begin{equation}
A_3 = E(t-t_0 ) \equiv A+Et \ .
\label{a3def2}
\end{equation}
In this case, one does have spatial translational invariance, but there
is no invariance under arbitrary temporal shifts. A translation in time
corresponds to a shift in the constant component $A$ of the gauge field,
and, on a space of finite extent, different $A$ are in general physically
inequivalent, since only gauge transformations which shift $A$ by certain
finite increments exist. Therefore, the Hamiltonian in the presence of the
field (\ref{a3def2}) is time-dependent\footnote{Note that one cannot argue
external fields of the type (\ref{a3def}) to be gauge-equivalent to
time-independent ones such as (\ref{a0def}) on a finite coordinate space;
the corresponding gauge transformation conflicts with the boundary
conditions.} (with a periodicity which depends on the strength of the
electric field $E)$.

These observations affect the above analysis in two ways. For one, the
correlator (\ref{2ptdecay}) really depends on two external parameters,
$E$ and $A$. As a consequence, the correlator ratio (\ref{ratio2})
contains all quadratic dependences which can result in the presence of
the two parameters,
\begin{equation}
R_2 (t) \longrightarrow \frac{1}{W_0 } \left( W_2^{AA} A^2 + W_2^{AE} AE
+ W_2^{EE} E^2 \right) - \left( m_2^{AA} A^2 + m_2^{AE} AE + m_2^{EE} E^2
\right) t \ .
\label{ratioAE}
\end{equation}
On the other hand, for a small external field, the ground-state neutron
mass and wave function adjust adiabatically as time evolves; this implies
that the overlap coefficients $W_2^{**} $ (and also $m_2^{**} $) in
(\ref{ratioAE}) are time-dependent, complicating the extraction of
the neutron mass shift from the slope of the correlator ratio as a
function of time. Even with the expanded set of data obtained within
the present work, not enough information is available to disentangle
these time dependences in the most general case. However, in one specific
situation, which will be argued below to be the relevant one as far as
the extraction of the neutron electric polarizability is concerned, the
slope of $R_2 (t)$ indeed does yield the neutron mass shift directly;
namely, when the Hamiltonian is stationary in time. In that case,
time dependences in the coefficients $W_2^{**} $ (and $m_2^{**} $) are
relegated to higher than linear order\footnote{As already mentioned at
the end of section \ref{2ptsec}, and discussed further in section
\ref{massres}, here, an additional technical issue arises: While
stationarity of the Hamiltonian guarantees that the neutron wave
function is stationary, there is, in addition, a time dependence
contained in the smeared neutron sink via eqs.~(\ref{ords1}),(\ref{ords2}).
As a result, the overlap between neutron wave function and sink can still
contain contributions linear in time if one insists on manifest invariance
of the neutron sink with respect to gauge transformations of the external
gauge field, implying the inclusion of (\ref{ords1}),(\ref{ords2}) in the
smeared sink construction. On the other hand, if one restricts the
calculation to the fixed gauge field (\ref{a3def2}) and foregoes manifest
invariance of the neutron sink with respect to gauge transformations of the
external field, it is legitimate to use the time-independent smeared sink
(\ref{ords0}) alone. In terms of the diagrammatic representation of
Fig.~\ref{fig4pt}, this corresponds to discarding all diagrams involving
smeared sinks other than \parbox{1cm}{\Huge $\circ $} \hspace{-0.7cm}.
In the analysis below, both options will be treated, and the final
result for the neutron electric polarizability will be seen to be
uninfluenced by this choice. A way to avoid this issue, not explored
within the present investigation, would be to use a point neutron sink;
such a sink would be simultaneously time-independent and invariant under
gauge transformations of the external field. On the other hand, a point
sink would have a small overlap with the true neutron wave function,
implying a lessened efficiency in the extraction of the neutron ground
state signal.}, and one can indeed equate, up to a minus sign, the slope
of $R_2 (t)$, cf.~(\ref{ratioAE}), with the mass shift
\begin{equation}
\Delta m = m_2^{AA} A^2 + m_2^{AE} AE + m_2^{EE} E^2 \ .
\label{quadform}
\end{equation}
Moreover, since a shift in $A$ is equivalent to a shift in time,
stationarity of the Hamiltonian in time also implies stationarity in
$A$, i.e., the mass shift (\ref{quadform}) (and consequently the slope
of $R_2 (t)$) is stationary in $A$ in this particular situation. Thus,
in analyzing the measured data below, the slope of $R_2 (t)$,
\begin{equation}
S_2 = \frac{dR_2 }{dt} \ ,
\label{s2def}
\end{equation}
will be extracted\footnote{In practice, the average slope over a fixed
measurement time interval will be determined in order to reduce the
statistical uncertainty.} and, for given $E$, the unique external
field will be sought out at which $S_2 $ is stationary with respect
to $A$. Since this is then necessarily the point at which the
Hamiltonian is stationary in time, at that point, then, one can
identify
\begin{equation}
S_2 = -\Delta m \ .
\label{s2id}
\end{equation}
To complete the analysis, it is necessary to discuss in more detail the
dependence of the neutron mass shift $\Delta m$ on the parameters $E$ and
$A$, and, in particular, the relevance of stationarity in $A$. In general,
the part of the neutron mass shift which is of second order in the
external gauge field (\ref{a3def2}) can be written as a quadratic form
in the parameters $E$ and $A$, cf.~(\ref{quadform}). However, this
quadratic form is not yet defined unambiguously and its coefficients
can consequently not yet all be interpreted as bona fide physical
properties of the neutron. To see this, consider shifting the entire
neutron mass measurement process by a time increment $\bar{t} $, i.e.,
the neutron source, which starting with (\ref{2ptdecay}) has so far been
assumed to be located at $t=0$, shall, for the sake of the following
argument, now be located at $t=\bar{t} $. If one concomitantly introduces
a shifted time coordinate $t^{\prime } = t-\bar{t} $ and a shifted
\begin{equation}
\bar{A} = A+E\bar{t} \ ,
\label{bara}
\end{equation}
then, in terms of the shifted quantities, the problem takes a form
identical to the original one, i.e., one measures the mass shift
\begin{eqnarray}
\Delta m &=& m_2^{AA} \bar{A}^{2} + m_2^{AE} \bar{A} E + m_2^{EE} E^2
\label{shifta} \\
&=& m_2^{AA} A^2 + (m_2^{AE} +2\bar{t} m_2^{AA} ) AE
+(m_2^{EE} +\bar{t} m_2^{AE} + \bar{t}^{\, 2} m_2^{AA} ) E^2
\label{bartra} \\
&=& \bar{m}_{2}^{AA} A^2 + \bar{m}_{2}^{AE} AE + \bar{m}_{2}^{EE} E^2 \ .
\label{barco}
\end{eqnarray}
Thus, in terms of the original definition of $E$ and $A$, cf.~(\ref{a3def2}),
the shifted measurement yields a quadratic form for the mass shift with
different coefficients $\bar{m}_{2}^{AE} $ and $\bar{m}_{2}^{EE} $ (whereas
the remaining coefficient is invariant, $\bar{m}_{2}^{AA} =m_2^{AA} $).
Therefore, the question arises how the neutron electric polarizability
is to be extracted from the total mass shift $\Delta m$; evidently,
polarizability effects enter both the coefficients $\bar{m}_{2}^{EE} $
and $\bar{m}_{2}^{AE} $, which can be traded off against one another, as
demonstrated above.

As a first step towards disentangling the different effects at play, the
immutable character of the coefficient $\bar{m}_{2}^{AA} $ should be
noted, which allows it to be interpreted as an unambiguous property of
the neutron. This property moreover is separate from the electric
polarizability; as verified by explicit calculation below, also at
$E=0$ one obtains the mass shift $\Delta m = \bar{m}_{2}^{AA} A^2 $,
encoding the response of the neutron to distortion by the presence of
the constant background field\footnote{Note that the effect of such a
constant background field is equivalent to a modification of the boundary
conditions in the relevant direction, introducing nontrivial Bloch momenta
varying with quark flavor.}. With $\Delta m = \bar{m}_{2}^{AA} A^2 $
representing the response of a neutron already in the absence of any
external electric field, one would indeed expect modifications of this
response due to a distortion of the neutron by an additional electric
field to occur only at higher than quadratic order in the external
gauge field. The representation-independence of $\bar{m}_{2}^{AA} $,
i.e., its independence of the choice of $\bar{t} $, thus seems plausible,
and is consistent with the interpretation of the $\bar{m}_{2}^{AA} A^2 $
term as a response separate from the electric polarizability.

On the other hand, in view of (\ref{bartra}), there is one special
representation, i.e., choice of $\bar{t} $, which seems particularly
transparent, namely, the representation in which $\bar{m}_{2}^{AE} $
vanishes, such that\footnote{A way to understand how the simplified
dependence (\ref{barqf}) arises is the following: Choosing $\bar{t} $
such as to realize (\ref{barqf}) shifts the mass shift measurement time
interval towards the time $t=0$. Now, in view of the definition
(\ref{a3def2}), the time $t=0$ is special in that the $A$- and
$E$-directions in parameter space are, in a sense, orthogonal there:
At $t=0$, a change of $E$ affects only the slope of $A_3 $, but not its
value; at other times, this is not the case and a change in $E$ also
implies an adjustment of the value of $A_3 $ itself, which could be
equally effected (or compensated) by a change in $A$. It is this
implicit relation between $E$ and $A$ which generates the coupled
dependence (\ref{quadform}); however, if one measures near $t=0$,
the implicit relation is dissolved and it is natural to obtain
the decoupled dependence given by (\ref{barqf}).}
\begin{equation}
\Delta m = \bar{m}_{2}^{AA} A^2 + \bar{m}_{2}^{EE} E^2 \ .
\label{barqf}
\end{equation}
The form (\ref{barqf}) suggests an interpretation of the data in terms of
two, now cleanly disentangled effects, namely, the polarizability effect
determined by $\bar{m}_{2}^{EE} $ and the effect of introducing a constant
background field, embodied in the coefficient $\bar{m}_{2}^{AA} $. Thus,
in this special representation, the electric polarizability is given by
\begin{equation}
\alpha = -2\bar{m}_{2}^{EE} \ .
\label{alphaEE}
\end{equation}
In other words, to isolate the electric polarizability effect from the
complete mass shift, one simply sets $A=0$ in the representation
(\ref{barqf}).

Finally, it is possible to rephrase this prescription for extracting
the neutron electric polarizability in a manner which is independent of
the particular representation, i.e., the choice of $\bar{t} $. Setting
$A=0$ in the representation (\ref{barqf}) is tantamount to evaluating
the mass shift $\Delta m$ at the extremum in $A$. However, this way of
stating the prescription does not rely on that specific representation;
after all, in view of (\ref{shifta}) and (\ref{barco}) in conjunction
with (\ref{bara}), different representations are related by shifting
the value of $A$, and the extremum of $\Delta m$ as a function of $A$
is invariant under such shifts. Thus, one can isolate the neutron
electric polarizability in any and all representations by seeking out
the stationary point of the mass shift as a function of $A$.

In view of this, and the equivalence of shifts in $A$ with shifts in time,
the neutron electric polarizability can indeed be extracted by considering
the correlator ratio $R_2 (t)$, cf.~(\ref{ratioAE}), specifically for
external gauge fields in the vicinity of which the Hamiltonian is
stationary in time; this validates the arguments presented further
above in conjunction with eqs.~(\ref{quadform})-(\ref{s2id}).

\subsection{Spatially varying gauge field}
\label{a0sec}
In the case of the gauge field (\ref{a0def}),
\begin{equation}
A_0 = -Ex_3 \ ,
\label{a0def2}
\end{equation}
one does have a time-independent Hamiltonian, and consequently one can
straightforwardly extract the energy of the neutron ground state from the
exponential time decay of the neutron two-point function. However, this
invariance under translations in time comes at the expense of breaking
spatial translational invariance. The linear dependence of (\ref{a0def2})
on $x_3 $ conflicts with the periodic boundary conditions; when traveling
through the lattice in the 3-direction, as the boundary is traversed,
$A_0 $ is forced to jump, implying a spike in the electric field which
is present in addition to the constant electric field induced by
(\ref{a0def2}).

Therefore, the neutron is not propagating in a spatially homogeneous
background and its momentum is not a good quantum number\footnote{It is,
of course, still legitimate to use a zero-momentum neutron sink, as in
(\ref{2ptdecay}), since it will presumably have a finite overlap with
the true neutron ground state wave function. However, that wave function
itself will not carry a definite momentum.}. As a result, the ground state
energy one extracts from the decay of the neutron two-point function
contains not only the desired mass shift associated with the electric
polarizability, but further contributions due to, e.g., the effective
movement in a spatially varying potential and additional distortions
of the neutron by the electric field spikes. Within the present
investigation, no prescription for disentangling the neutron electric
polarizability from these other effects with a level of cogency
comparable to the one discussed in the previous section emerged.
Nevertheless, the ground state energy obtained below using the
external gauge field (\ref{a0def2}) is consistent with the polarizability
mass shift obtained using the external field (\ref{a3def2}), suggesting
that the contamination by the additional effects mentioned above is not
dominant. At least as far as the neutron's effective propagation in a
spatially varying potential is concerned, this seems plausible, since
the quantum mechanical zero-point energy associated with such motion
is suppressed by the comparatively large mass of the neutron.

Note that, although superficially the external fields (\ref{a3def2}) and
(\ref{a0def2}) seem quite similar, and simply related by an exchange of
the temporal with a spatial direction, the physical issues arising in the
two cases are quite distinct. This is due to the way the mass measurement
is set up. Up to exponentially suppressed effects, the neutron mass is
determined by physics within a limited time interval, between neutron
source and sink. The temporal boundaries, located far in the past or the
future of the measurement, have a negligible effect on the latter. By
contrast, one cannot similarly contain the region relevant for the
measurement in the spatial directions. In situations with spatial
translational invariance, by projecting onto a definite momentum, one
explicitly weights all of space equally during the entire measurement
process. Even in the absence of spatial translational invariance, it is
up to the dynamics to determine whether there is a significant probability
of finding the neutron near the spatial boundary. Thus, in general, the
spatial boundary conditions have a crucial influence on the problem.

In the case of the external field (\ref{a3def2}) discussed in the previous
section, this entails that shifts of the gauge field $A_3 $ by a constant
$A$ have a physical effect, since gauge transformations designed to remove
such a shift conflict with the spatial boundary conditions. As a
consequence, physics varies locally with time, as discussed extensively
further above. On the other hand, the neutron mass measurement is
insensitive to the behavior of the external field at the temporal
boundaries.

In the case of the external field (\ref{a0def2}), one encounters a largely
converse situation: The neutron ground state is sensitive to the spatial
boundary, at which it encounters spikes in the external electric field;
on the other hand, as long as one is not in the vicinity of the boundary,
physics does not vary locally in space. The latter is due to the fact
that one can indeed remove constant shifts in the field $A_0 $ {\em in
the time interval relevant for the neutron mass measurement} using gauge
transformations. These transformations do of course need to exhibit
additional nontrivial structures located far in the past and the future
of the measurement, but these structures will not influence the measurement.
This also motivates the fact that no explicit freedom of shifting $A_0 $
by a constant is included in (\ref{a0def2}), in contradistinction to
(\ref{a3def2}). Such shifts are not expected to yield new physics according
to the above argument.

Comparing the two cases, ultimately (\ref{a3def2}) can be treated in a more
satisfactory fashion because the positioning of the neutron source and sink
allows one to contain and control the breaking of temporal translational
invariance introduced by  the field (\ref{a3def2}). By contrast, in the case
of  (\ref{a0def2}), there is no analogous control; the neutron dynamics
must be allowed to explore space and, in general, the breaking of spatial
translational invariance will influence the measurement in a nontrivial
fashion.

\section{Measurement results}
\subsection{Quark wave function renormalization}
\label{rensec}
\begin{figure}[t]
\centerline{\epsfig{file=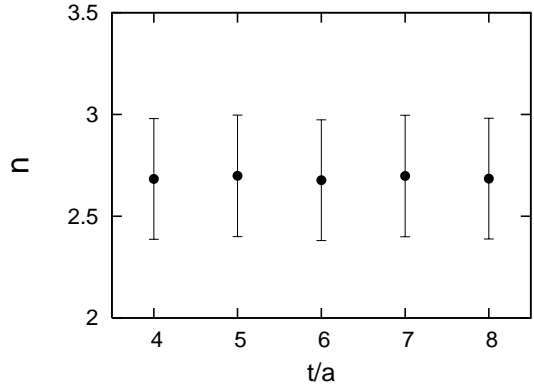,width=5.3cm,angle=-90} }
\caption{(Unrenormalized) number of valence quarks $n$ in the neutron
measured for a range of insertion times. The neutron source is located
at $t=0$ and the neutron sink at $t=13a$.}
\label{plateau}
\end{figure}

To determine the renormalization factor $z_V $ in (\ref{se1}),(\ref{se2}),
a measurement of the number of valence quarks in the neutron was carried
out and subjected to the condition that this number equal three. In
practice, this is realized by measuring the appropriate three-point
function, i.e., a diagram of the type $I03$, with the difference that
the lone operator insertion is of the form of $M_1 $, cf.~(\ref{se1}),
without the weighting by the quark electric charge $q_f $, and with a
formal external gauge field
\begin{equation}
A_0 (x) = \delta (x_0 -t) \ ,
\label{adelta}
\end{equation}
where $t$ is a time between neutron source and sink. Normalizing this by
the neutron two-point function yields, up to an additional factor
$i$ stemming from the Euclidean treatment of the time coordinate, the
(lattice analogue of the) expectation value of $\int d^4 x\, j_0 A_0 $
in the neutron, where $j_0 $ denotes the temporal component of the quark
current. In view of (\ref{adelta}), this reduces to the number of (valence)
quarks $n=\int d^3 x\, j_0 $ present at the time $t$. Fig.~\ref{plateau}
displays the plateau obtained measuring $n$ at different insertion times
$t$ for fixed neutron source and sink. Taking the average of the displayed
plateau values, one infers
\begin{equation}
z_V = 1.12 \pm 0.12 \ ,
\end{equation}
where the uncertainty was obtained using the jackknife method. This
measurement of $z_V $ enters all further measurements below; its
uncertainty will be jackknifed into those measurements.

\subsection{Neutron mass shift}
\label{massres}
Measurements of the neutron mass shift according to the discussion
in section \ref{interpsec} were taken for the following cases: With the
location of the neutron source once again defining $t=0$, the external
gauge field (\ref{a3def2}) was studied for the cases $t_0 = -10a$,
$t_0 = 0$ and $t_0 = 6a$ (where $a$ denotes the lattice spacing) at a
fixed nonvanishing value of $E$. Furthermore, the case $E=0$ at a fixed
nonvanishing value of $A$ in (\ref{a3def2}) was investigated. The external
gauge field (\ref{a0def2}) was treated for a fixed nonvanishing value of
$E$, where the plane $x_3 =0$ was taken to define locations maximally
distant from the lattice boundary in the 3-direction, and was
simultaneously used as the location of the neutron source (i.e., the
smeared quark sources were constructed using an initial position $x$ in
(\ref{smearmat}) located in the $x_3 =0$ plane).

Furthermore, as discussed in section \ref{calcsec}, to suppress fluctuations
in disconnected diagrams, these external gauge fields were chopped in the
time direction, i.e., $A_3 \equiv 0$ and $A_0 \equiv 0$ for $t< -a$ and
$t>14a$ in the following. Only for $-a\leq t\leq 14a$ do $A_3 $ and $A_0 $
take the forms (\ref{a3def2}) and (\ref{a0def2}), respectively. The
temporal boundaries of the lattice, at which Dirichlet boundary conditions
are enforced on the quark fields, are located at $t=-10a$ and $t=22a$.
Stochastic estimation of the disconnected diagrams was based on 120
stochastic sources, as described in section \ref{calcsec}, except for
the cases $t_0 =-10a$ and $E=0$ in (\ref{a3def2}), for which 240
stochastic sources were used.

\begin{figure}[p]
\centerline{\hspace{0.2cm}
\epsfig{file=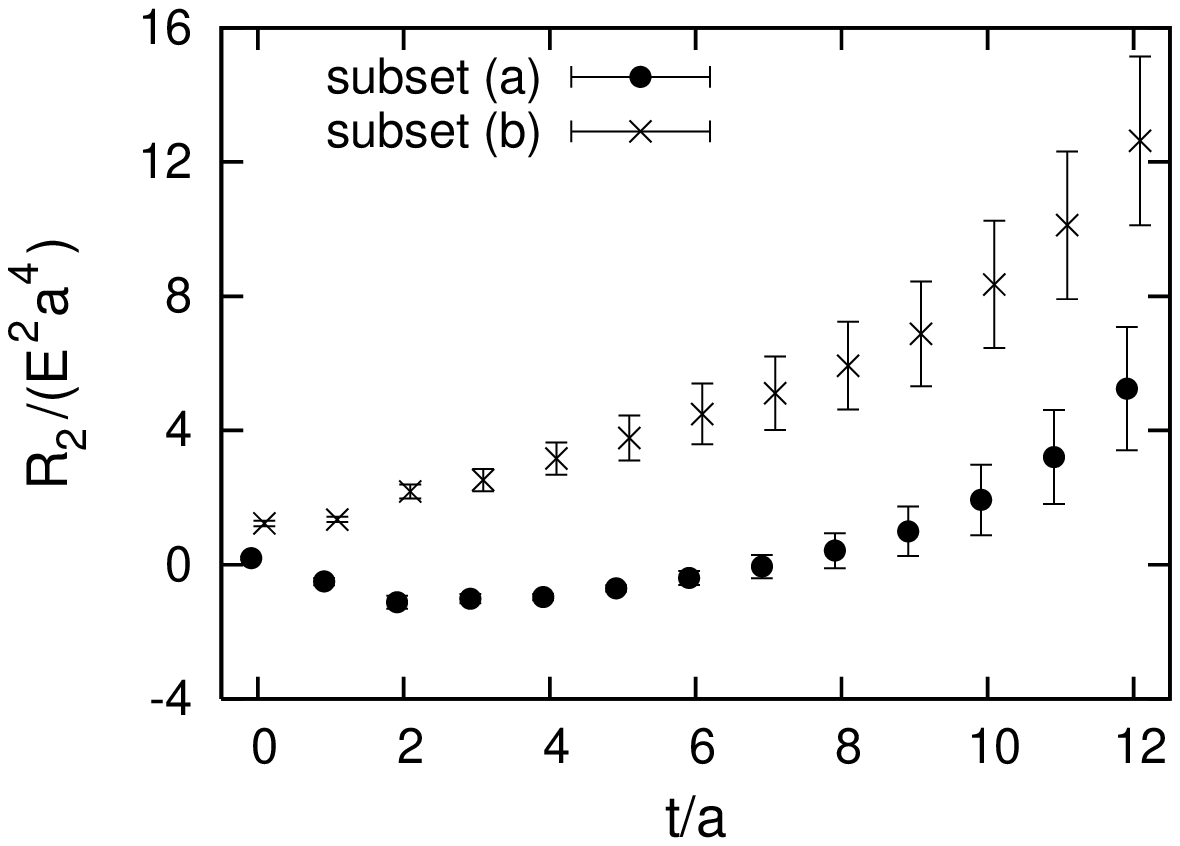,width=5.3cm,angle=-90}
\hspace{0.5cm}
\epsfig{file=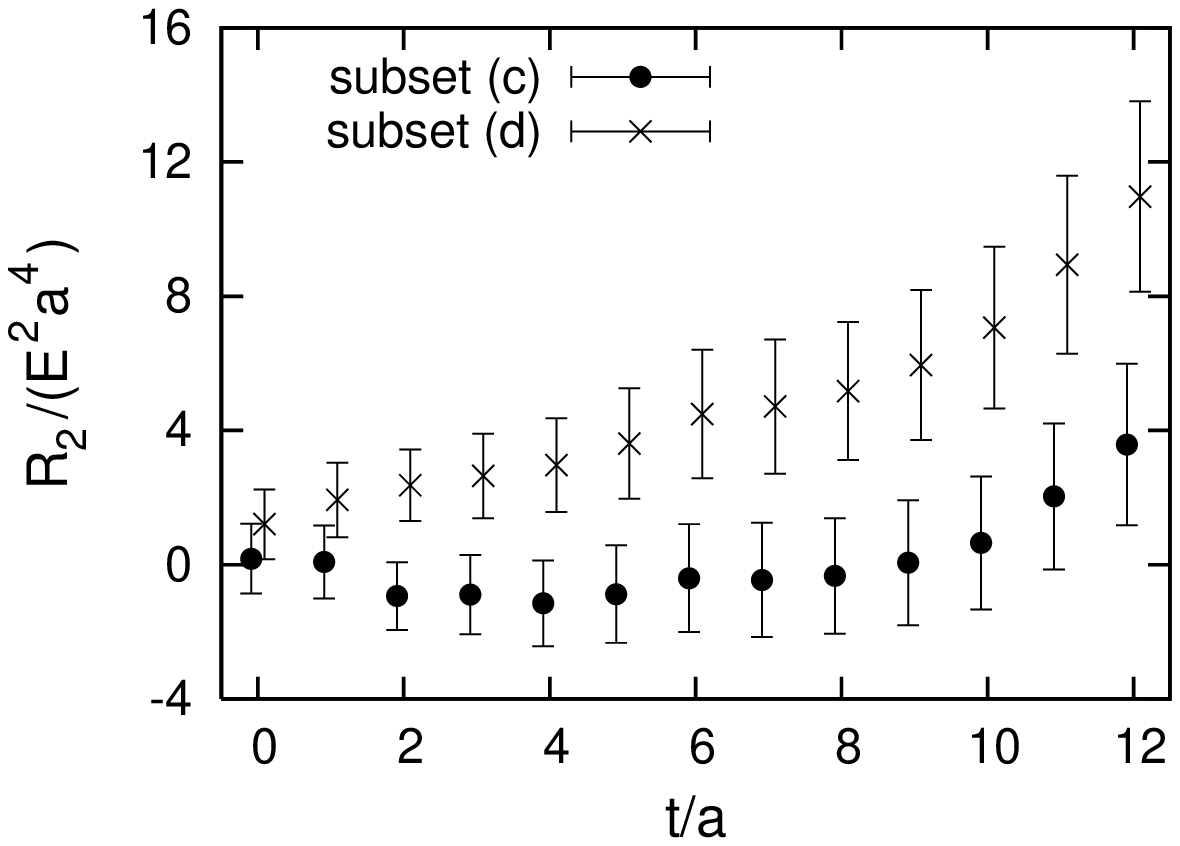,width=5.3cm,angle=-90} }
\caption{Contributions by selected subsets of diagrams, as specified
in the main text, to the ratio $R_2 $, as a function of temporal
source--sink separation $t$. All measurements are taken at
integer times; data are slightly displaced from those times in the
figures for better readability. These results were obtained using an
external field of the form (\ref{a3def2}), i.e., $A_3 =E(t-t_0 )$,
with $t_0 =-10a$, where $t=0$ corresponds to the neutron source location.
The electric field $E$ providing the scale is cast in Gau\ss ian units.}
\label{r2plot1}
\end{figure}

\begin{figure}[p]
\centerline{\hspace{0.2cm}
\epsfig{file=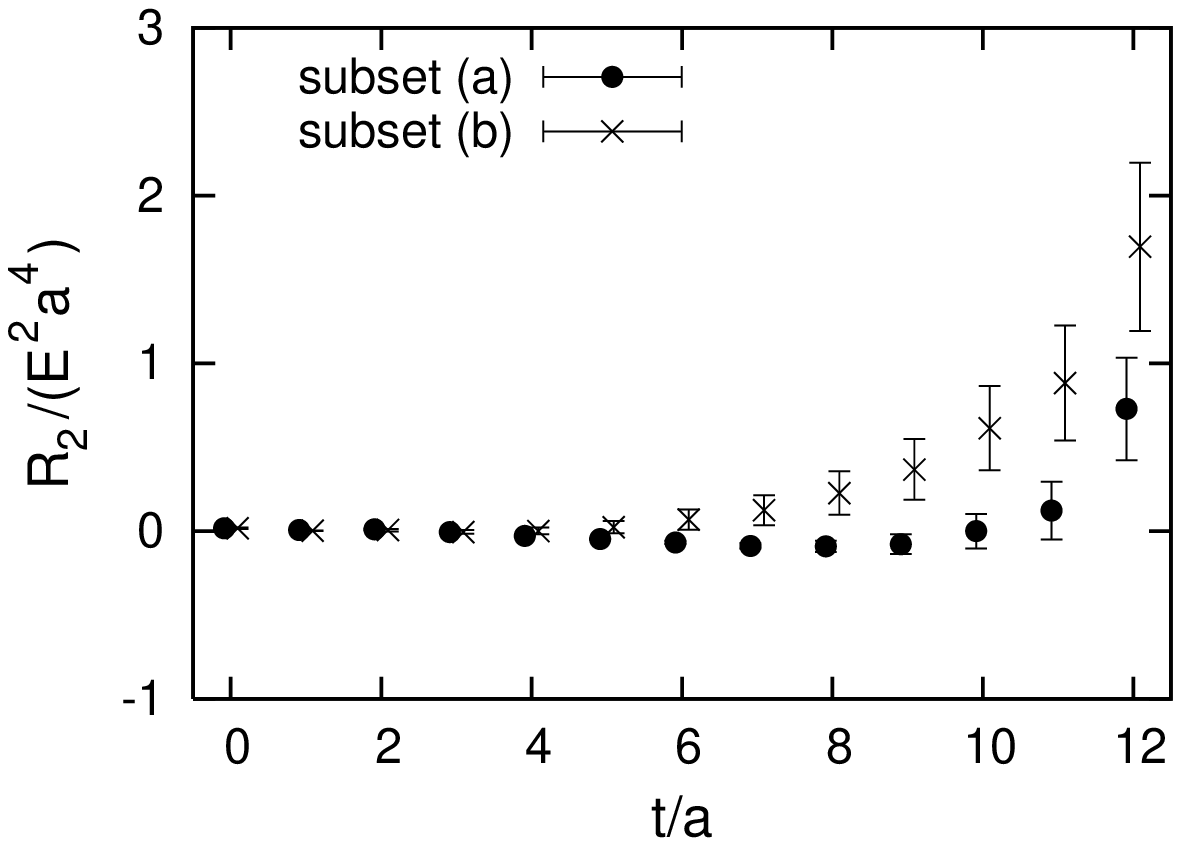,width=5.3cm,angle=-90}
\hspace{0.5cm}
\epsfig{file=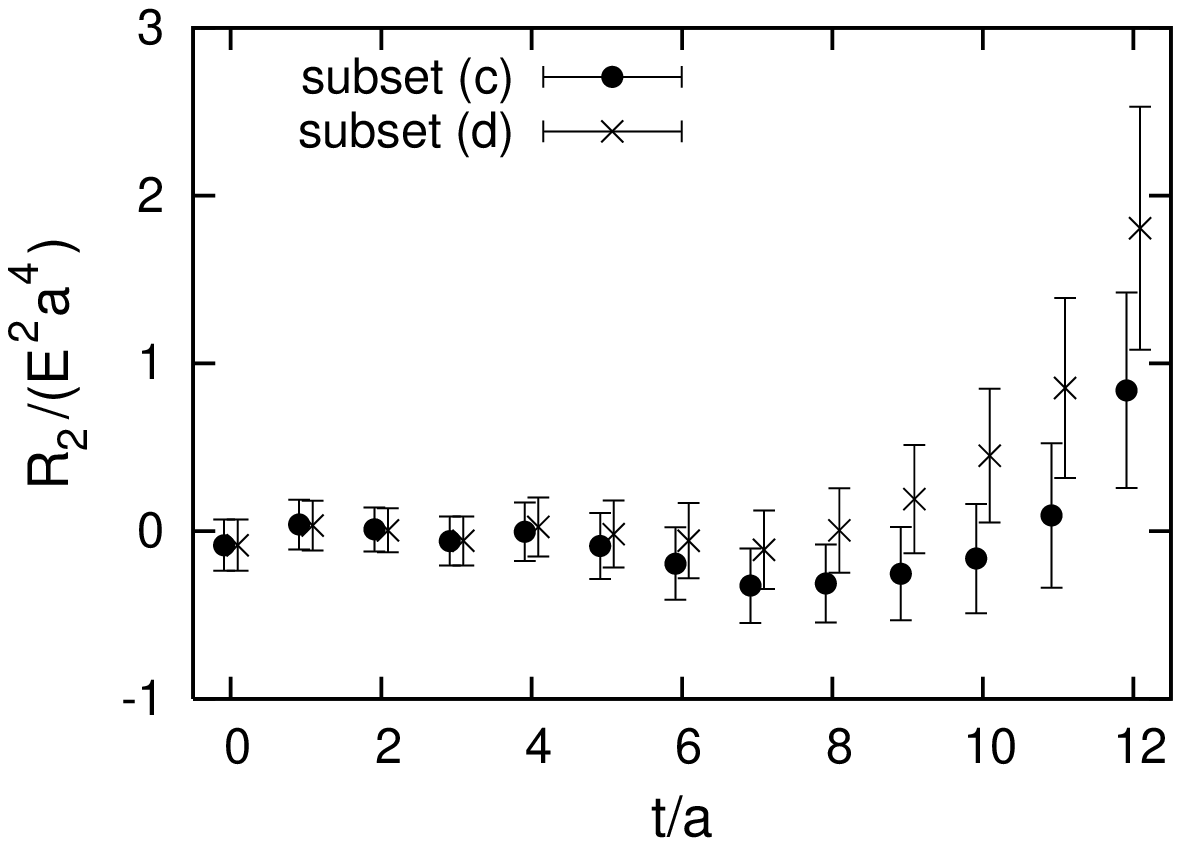,width=5.3cm,angle=-90} }
\caption{Contributions by selected subsets of diagrams, as specified
in the main text, to the ratio $R_2 $, as a function of temporal
source--sink separation $t$. All measurements are taken at
integer times; data are slightly displaced from those times in the
figures for better readability. These results were obtained using an
external field of the form (\ref{a3def2}), i.e., $A_3 =E(t-t_0 )$,
with $t_0 =0$, where $t=0$ corresponds to the neutron source location.
The electric field $E$ providing the scale is cast in Gau\ss ian units.}
\label{r2plot2}
\end{figure}

\begin{figure}[p]
\centerline{\hspace{0.2cm}
\epsfig{file=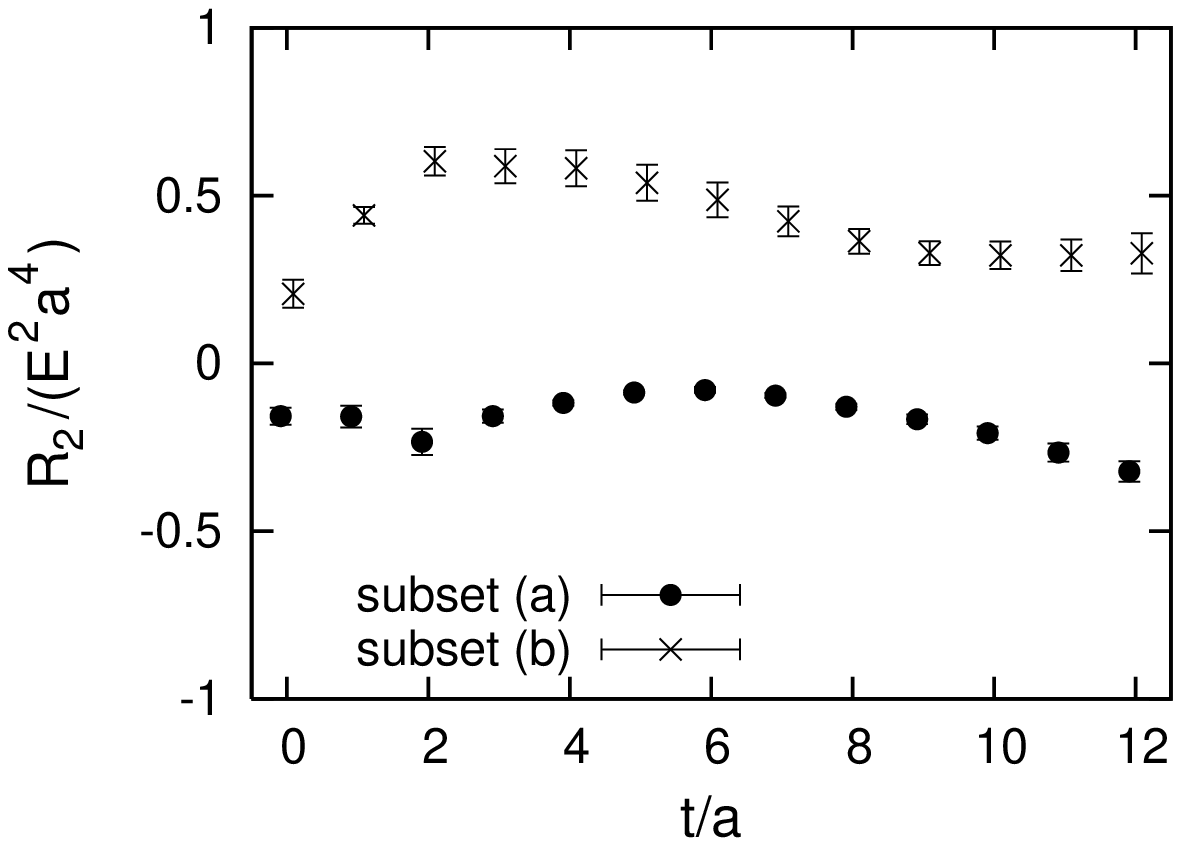,width=5.3cm,angle=-90}
\hspace{0.5cm}
\epsfig{file=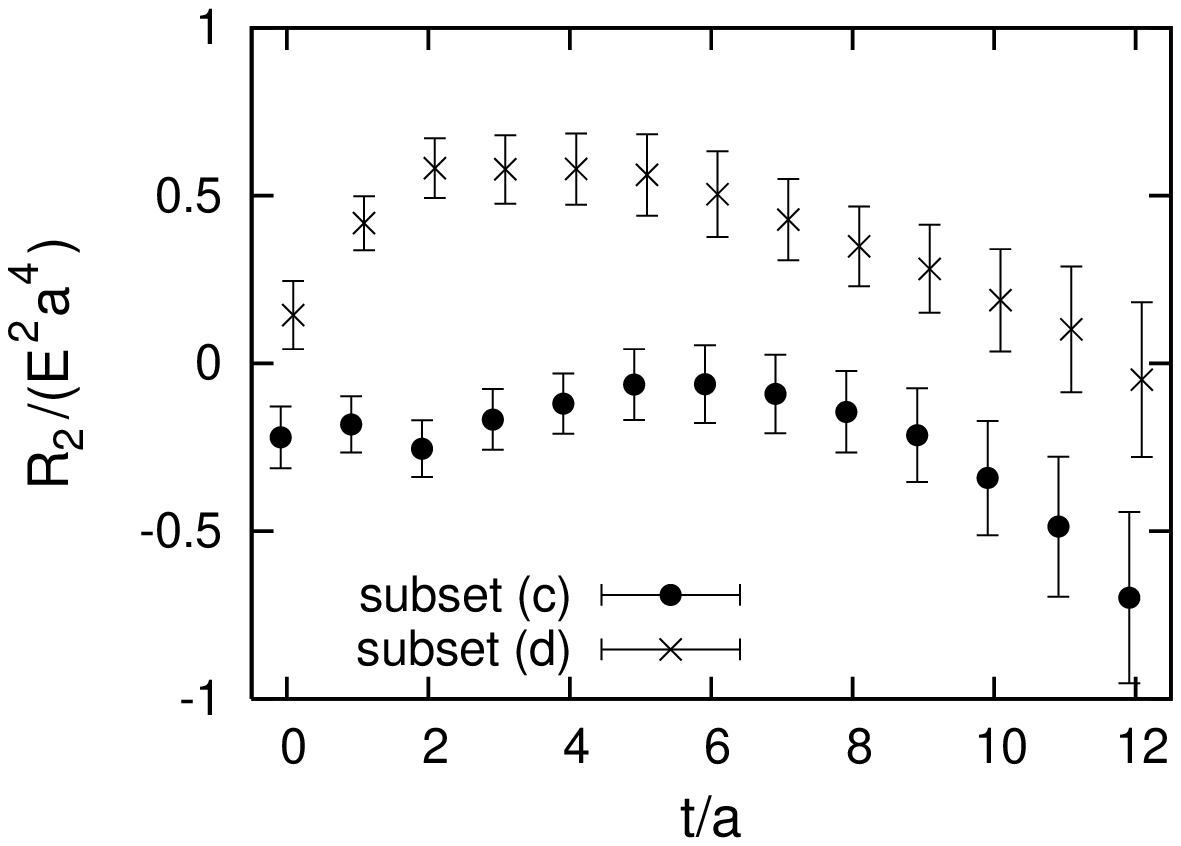,width=5.3cm,angle=-90} }
\caption{Contributions by selected subsets of diagrams, as specified
in the main text, to the ratio $R_2 $, as a function of temporal
source--sink separation $t$. All measurements are taken at
integer times; data are slightly displaced from those times in the
figures for better readability. These results were obtained using an
external field of the form (\ref{a3def2}), i.e., $A_3 =E(t-t_0 )$,
with $t_0 =6a$, where $t=0$ corresponds to the neutron source location.
The electric field $E$ providing the scale is cast in Gau\ss ian units.}
\label{r2plot3}
\end{figure}

\begin{figure}[p]
\centerline{\hspace{0.2cm}
\epsfig{file=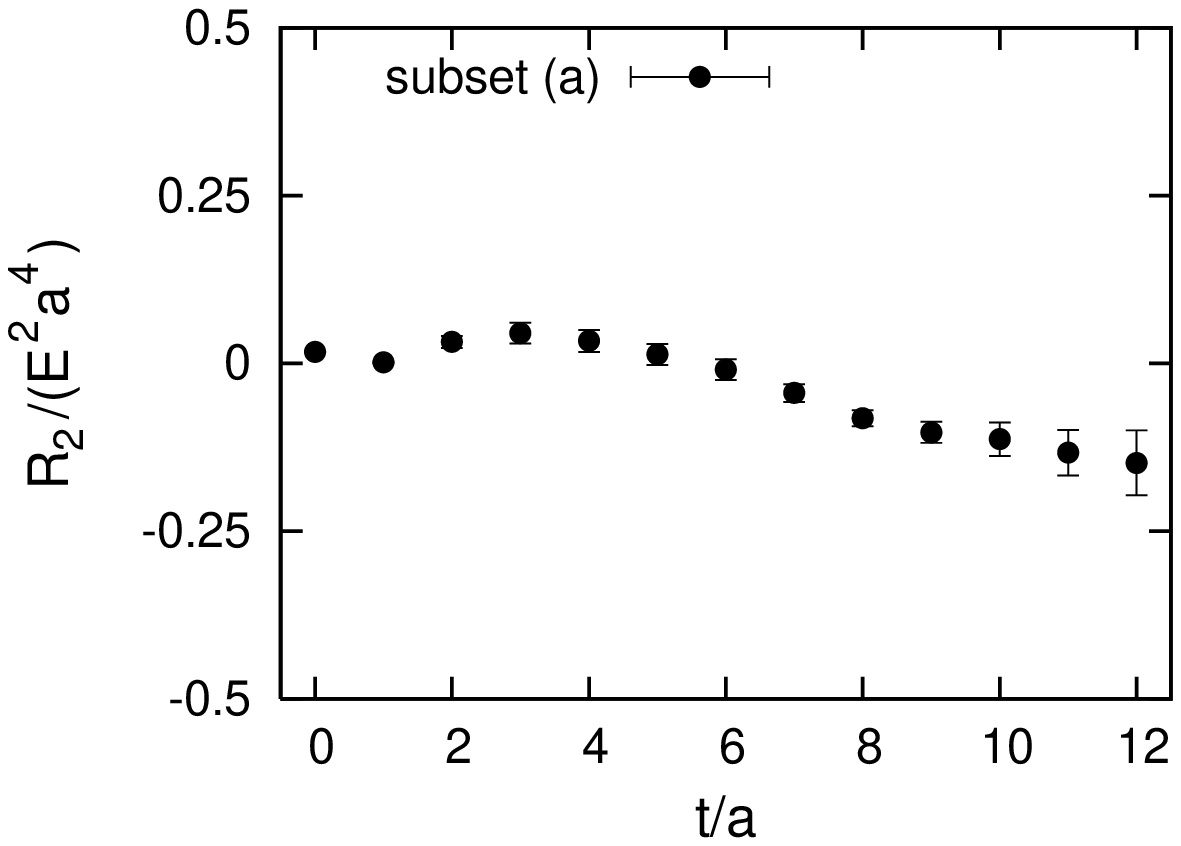,width=5.3cm,angle=-90}
\hspace{0.5cm}
\epsfig{file=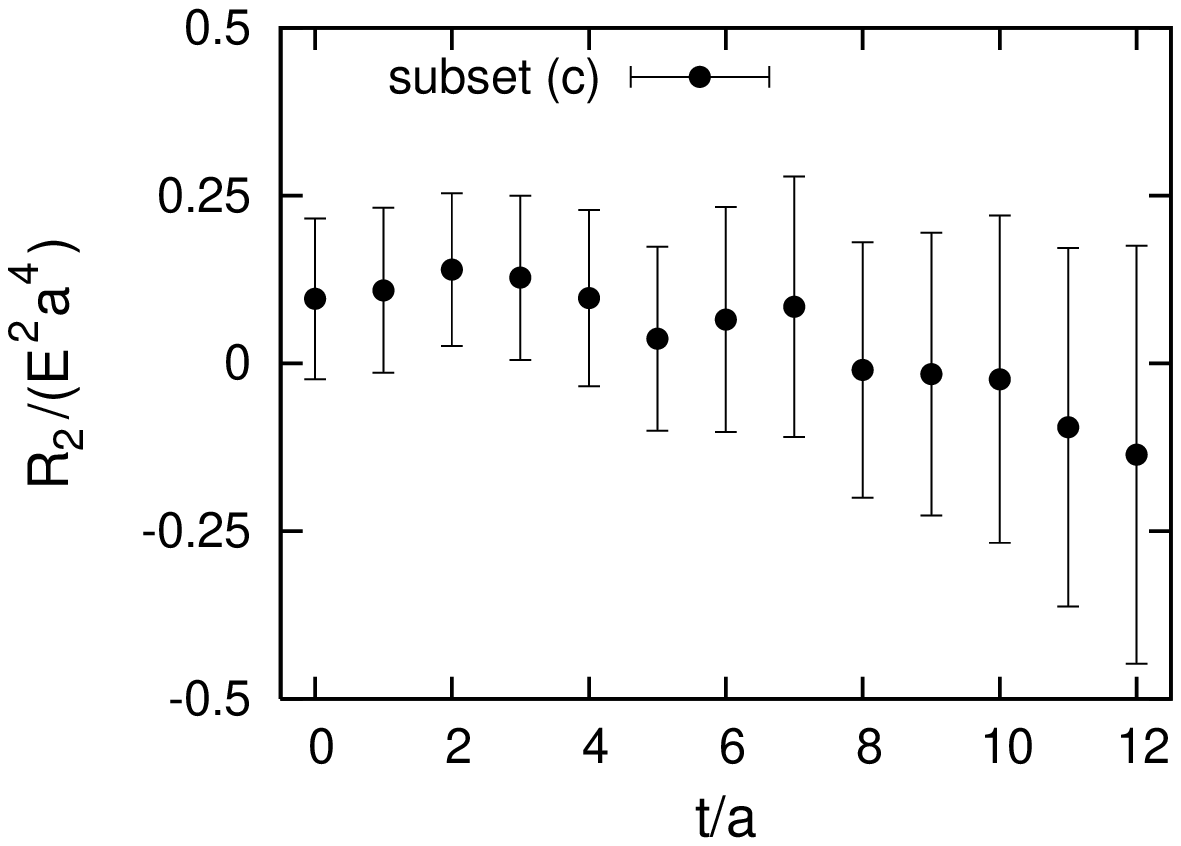,width=5.3cm,angle=-90} }
\caption{Contributions by selected subsets of diagrams, as specified
in the main text, to the ratio $R_2 $, as a function of temporal
source--sink separation $t$. These results were obtained using an
external field of the form (\ref{a0def2}), i.e., $A_0 =-Ex_3 $;
for this background, there are no smearing contributions beyond zeroth
order in the external field, i.e., case (b) is identical to case (a),
and case (d) is identical to case (c). The electric field $E$ providing
the scale is cast in Gau\ss ian units.}
\label{r2plot4}
\end{figure}

\begin{figure}[p]
\centerline{\hspace{0.2cm}
\epsfig{file=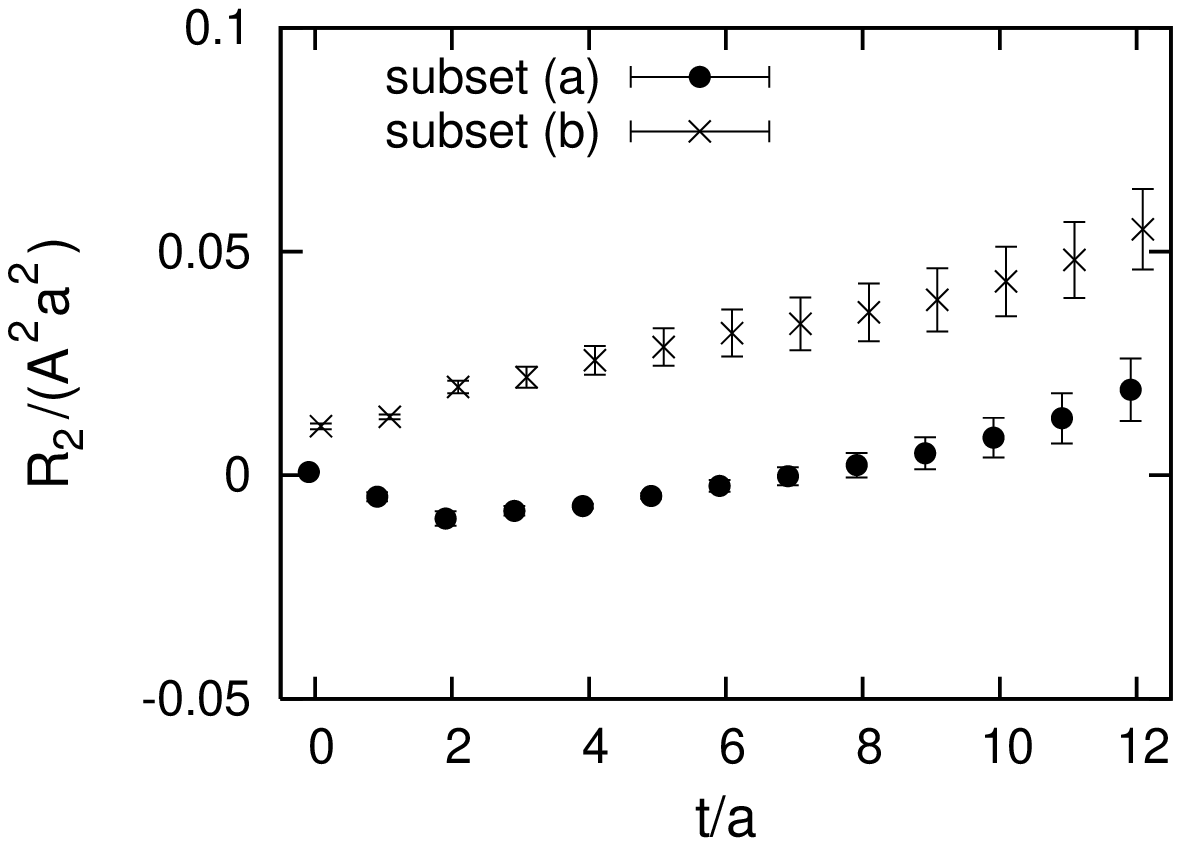,width=5.3cm,angle=-90}
\hspace{0.5cm}
\epsfig{file=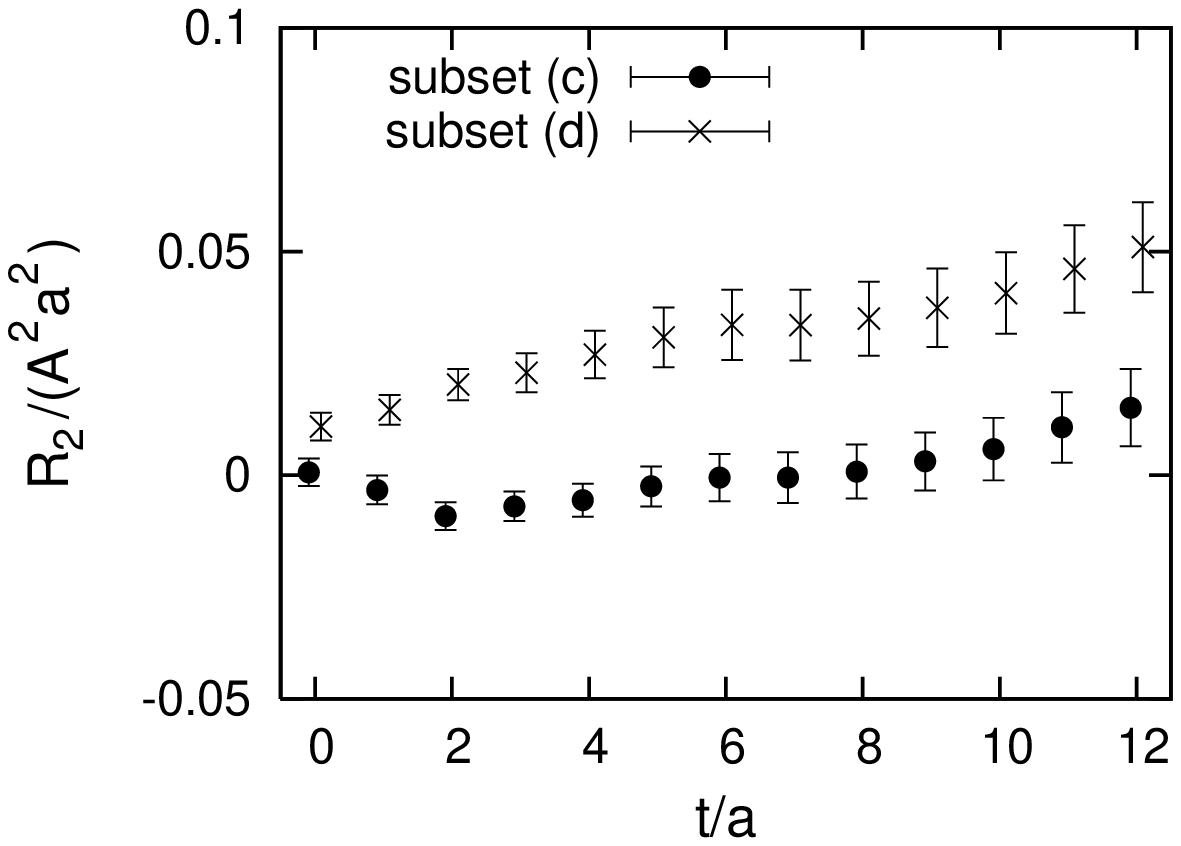,width=5.3cm,angle=-90} }
\caption{Contributions by selected subsets of diagrams, as specified
in the main text, to the ratio $R_2 $, as a function of temporal
source--sink separation $t$. All measurements are taken at
integer times; data are slightly displaced from those times in the
figures for better readability. These results were obtained using an
external field of the form (\ref{a3def2}) with $E=0$, i.e., $A_3 \equiv A$.
The constant background gauge field $A$ providing the scale is cast in
Gau\ss ian units.}
\label{r2plot5}
\end{figure}

Figs.~\ref{r2plot1}--\ref{r2plot5} display measurements of the ratio
\begin{equation}
R_2 (t) \equiv \frac{G^{(2)} (p=0,t)}{G_0 (p=0,t)} \ ,
\end{equation}
cf.~(\ref{ratio2}),(\ref{ratioAE}), for all the aforementioned external gauge
fields, in units of the relevant external field magnitude. That is, $R_2 $
is shown in units of $A^2 $ for the case $E=0$ and in units of $E^2 $ in the
other cases; furthermore, here and in the following, Gau\ss ian units are
adopted. Different subsets
of diagrams from Fig.~\ref{fig4pt} contributing to $R_2 $ are shown in the
individual plots (a)-(d) in each case. Figs.~\ref{r2plot1}--\ref{r2plot5} (a)
show only the contributions from connected diagrams with lowest-order
smearing, i.e., the diagrams $I0*$. Figs.~\ref{r2plot1}--\ref{r2plot5} (b)
show the result of including all connected diagrams, $I{**}$. Note that, in
the case of the external field (\ref{a0def2}), there are no connected
contributions beyond $I0*$, since smearing occurs only in the spatial
directions and thus never involves the gauge field component $A_0 $.
Figs.~\ref{r2plot1}--\ref{r2plot5} (c) show the result of including all
diagrams with lowest-order smearing, i.e., the diagrams $*0*$. Finally,
Figs.~\ref{r2plot1}--\ref{r2plot5} (d) show the sum of all diagrams depicted
in Fig.~\ref{fig4pt}. Note that, in the $SU(3)$ flavor-symmetric case
investigated in this work, the only nonvanishing disconnected diagrams
are $J01$ and $J03$, regardless of the external field used. This is due
to the fact that the disconnected loop with a linear external field
insertion is proportional to the sum of the quark charges, and therefore
vanishes. Thus, Figs.~\ref{r2plot1}--\ref{r2plot5} (c) and
Figs.~\ref{r2plot1}--\ref{r2plot5} (d) contain the same disconnected
contributions.

In comparing Figs.~\ref{r2plot1}--\ref{r2plot3}, which display the results
obtained using the external gauge field (\ref{a3def2}) for various $t_0 $,
the different vertical scales should be noted. Compared to the case
$t_0 =-10a$, the cases $t_0 =0$ and $t_0 =6a$ exhibit only very small
slopes, which are determined with relatively small uncertainties,
cf.~also Table~\ref{slopetab} below. The foremost observation to be drawn
from Figs.~\ref{r2plot1}--\ref{r2plot3} is that the slope of $R_2 $ indeed
depends sensitively on $t_0 $, or, equivalently, the constant offset
$A$ in the external gauge field (\ref{a3def2}), as expected. This is also
corroborated by the $E=0$ measurement displayed in Fig.~\ref{r2plot5}. As
explained in section \ref{a3sec}, these measurements taken together will
make it possible to disentangle the constant field effect from the electric
polarizability. Before proceeding towards this central goal, a few further
remarks about the data are in order.

\begin{figure}[p]
\centerline{\epsfig{file=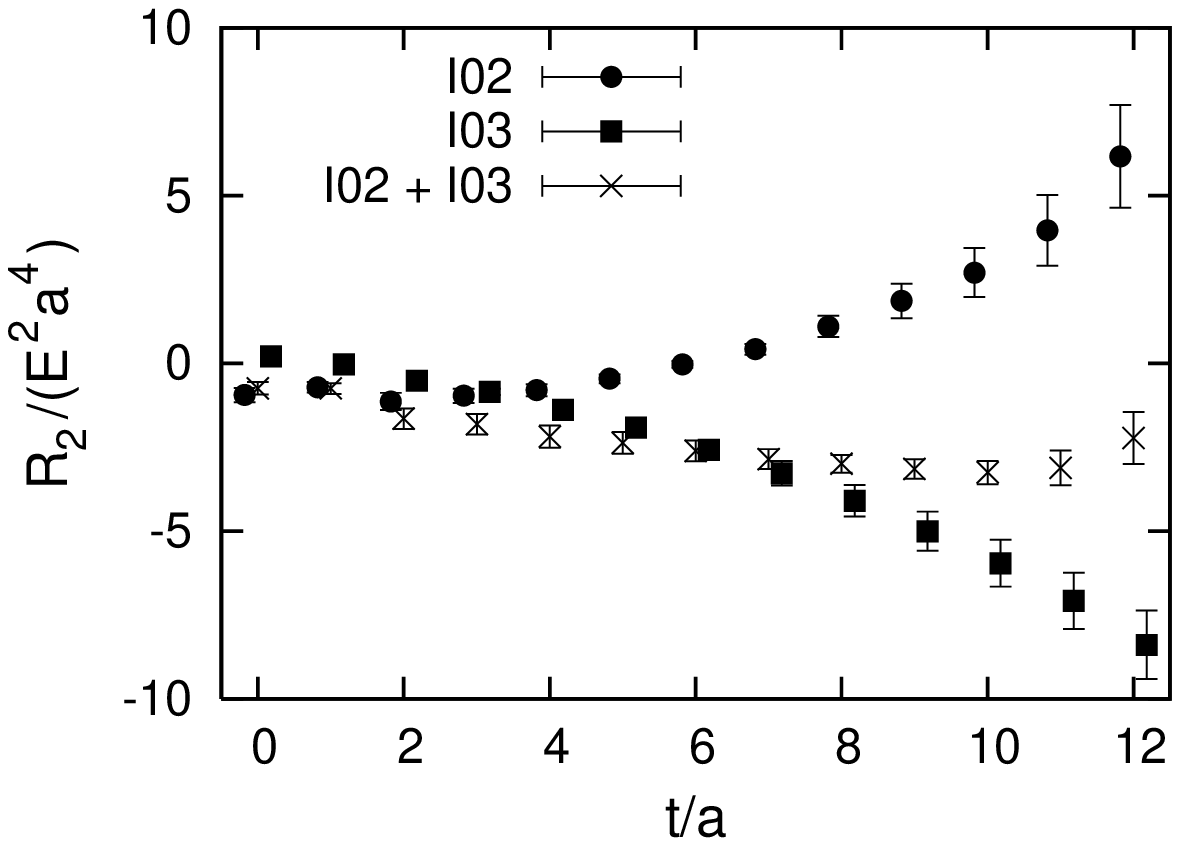,width=5.3cm,angle=-90} }
\caption{Contributions by diagrams $I02$ and $I03$, as well as their
sum, to the ratio $R_2 $, as a function of temporal source--sink
separation $t$. All measurements are taken at integer times; data are
slightly displaced from those times in the figure for better readability.
These results were obtained using an external field of the form
(\ref{a3def2}), i.e., $A_3 =E(t-t_0 )$, with $t_0 =-10a$, where $t=0$
corresponds to the neutron source location. The electric field $E$
providing the scale is cast in Gau\ss ian units.}
\label{plot2S}
\end{figure}

For one, there is a strong cancellation between the diagram $I02$ and the
corresponding contact term $I03$, cf.~Fig.~\ref{plot2S}. The contact term 
$I03$, which originates from expanding the gauge link variables to second
order in the external field, cf.~(\ref{4dcoup}),(\ref{vertic}),(\ref{se2}),
is not negligible, as a naive continuum limit might suggest; rather, it
contributes to the renormalization of the $I02$ diagram. Presumably, in
the continuum limit, it would be admissible to disregard diagrams such
as $I03$, at the expense of having to consider a strong renormalization
of the diagram $I02$ as its two vertices are permitted to approach each
other.

Secondly, one can furthermore observe from
Figs.~\ref{r2plot1}--\ref{r2plot5} that the disconnected diagrams
consistently tend to give a negative contribution to the slope
of the ratio $R_2 (t)$ for an external field of the form (\ref{a3def2}),
cf.~also Table~\ref{slopetab} below. For the external field (\ref{a0def2}),
the contribution is very slightly positive. However, it should be emphasized
that the contributions are in no case large enough to be significant compared
to the statistical uncertainty.

Thirdly, as noted in section \ref{a3sec}, the higher-order (in the external
field) sink smearing diagrams contained in Figs.~\ref{r2plot1}--\ref{r2plot3}
(b),(d) in general contribute additional linear time dependences to the ratio
$R_2 (t)$ beyond the ones associated with the mass shift of the neutron
in the external field. Indeed, a substantial difference in slope can be
seen\footnote{By considering the diagrams individually, one can indeed
verify that, as expected, the difference arises specifically due to sink
smearing contributions, and not source smearing contributions.} comparing
Fig.~\ref{r2plot1} (a) and (b), or also Fig.~\ref{r2plot1} (c) and (d).
To isolate the slope due to the mass shift itself, one should evaluate
Figs.~\ref{r2plot1}--\ref{r2plot3} (c); this comes at the expense of
foregoing a form of the neutron sink (and source) which is manifestly
invariant under gauge transformations of the external field. Of course, it
is not imperative to use a manifestly invariant form; all that is implied
by not doing so is that the neutron mass shift, a gauge-invariant
quantity, has been evaluated in a specific gauge for the external
electromagnetic field. Note, moreover, that this point is largely moot
at any rate, since the final result for the neutron electric polarizability
will be seen to not be affected significantly by the differences in
intermediate data introduced by the aforementioned sink smearing effects.

Returning to the main objective, extracting the neutron mass shift from
the slope of the ratio $R_2 (t)$, Table~\ref{slopetab} lists the slopes
extracted from the data displayed in Figs.~\ref{r2plot1}--\ref{r2plot5}.
These slopes were obtained by performing least-square fits of linear
functions in $t$ to the $R_2 (t)$ data for a range of $t$; the
uncertainties were obtained by jackknife analysis. The time range
used was $4a\leq t\leq 10a$; this choice minimizes statistical
uncertainty (by using as large a time range as possible) while still
allowing for a good least-square fit by a linear form as well as a
jackknife bias estimate small compared to the statistical
uncertainty\footnote{In the case of the external field
(\ref{a3def2}) with $t_0 = 0$ and $t_0 = 6a$, a more restricted
time range improves the linear fit, as is apparent, e.g., from
Fig.~\ref{r2plot3}; however, as already noted above, these two instances
are determined with very little uncertainty (regardless of the time range
used) compared to $t_0 = -10a$. As a result, it is the latter case which
dictates the choice of time range, which then was adopted for all cases
for consistency.}.

\begin{table}
\begin{center}
\begin{tabular}{|c||c|c|c|c|}
\hline
& (a) & (b) & (c) & (d) \\
\hline \hline
\parbox{2.8cm}{\mbox{\hspace{0.2cm} }
$S_2 /(a^3 E^2 )$ \\ (\ref{a3def2}), $t_0 = -10a$ } &
$0.46(18)$ & $0.83(24)$ & $0.26(26)$ & $0.63(30)$ \\
\hline
\parbox{2.5cm}{\mbox{\hspace{0.05cm} }
$S_2 /(a^3 E^2 )$ \\ (\ref{a3def2}), $t_0 =0$ } &
$0.000(16)$ & $0.096(38)$ & $-0.033(43)$ & $0.063(54)$ \\
\hline
\parbox{2.5cm}{\mbox{\hspace{0.05cm} }
$S_2 /(a^3 E^2 )$ \\ (\ref{a3def2}), $t_0 =6a$ } &
$-0.017(3)$ & $-0.047(7)$ & $-0.037(25)$ & $-0.067(27)$ \\
\hline
\parbox{2.5cm}{$-\Delta m /(a^3 E^2 )$ \\ \mbox{\hspace{0.6cm} }
(\ref{a0def2})} & $-0.027(4)$ & $-0.027(4)$ & $-0.019(41)$ & $-0.019(41)$ \\
\hline \hline
\parbox{2.5cm}{$-\Delta m /(aA^2 )$ \\ (\ref{a3def2}), $E=0$ } &
$0.0025(8)$ & $0.0028(8)$ & $0.0017(9)$ & $0.0020(10)$ \\
\hline
\end{tabular}
\end{center}
\caption{Slope $S_2 $, cf.~(\ref{s2def}), for different external
electromagnetic fields, in the appropriate external field units
and units of the lattice spacing $a$. Cases (a)-(d) correspond to
the different subsets of diagrams included in the corresponding
Figs.~\ref{r2plot1}-\ref{r2plot5} (a)-(d), cf.~main text.
In the case of the background field (\ref{a0def2}), as well as for
(\ref{a3def2}) with $E=0$, the slope $S_2 $ can be directly identified
with the negative mass shift, $-\Delta m$; hence the labeling of the
last two lines. By contrast, the data in the first three lines must be
processed further to locate the stationary point as a function of $t_0 $,
at which then $S_2 = -\Delta m$ can be identified, cf.~main text and
Table \ref{paratab}.}
\label{slopetab}
\end{table}

According to the discussion in section \ref{a3sec}, to extract the
electric polarizability from the data in the first three lines of
Table~\ref{slopetab}, these data should be viewed as defining a parabola
in $t_0 $, and the extremum in $t_0 $ should be sought out (note that
this is equivalent to viewing the data as defining a parabola in $A$,
since $A=-Et_0 $ and $E$ is constant). At the extremum, one can then
identify the slope $S_2 $ with the (negative) neutron mass shift, 
$-\Delta m$. Fitting the form (cf.~(\ref{quadform})-(\ref{s2id}))
\begin{eqnarray}
S_2 /E^2 &=&
-\left( m_2^{AA} A^2 + m_2^{AE} AE + m_2^{EE} E^2 \right) /E^2 \nonumber \\
&=& -\left( m_2^{AA} t_0^2 - m_2^{AE} t_0 + m_2^{EE} \right)
\label{parafit}
\end{eqnarray}
to the data in Table~\ref{slopetab} yields parabolas with the extrema and
curvatures listed in Table~\ref{paratab}.
The uncertainties quoted in Table~\ref{paratab} were again obtained using
the jackknife method. The quadratic coefficient $m_2^{AA} $ extracted
in this way agrees well with the $E=0$ values\footnote{The last lines
of Tables~\ref{slopetab} and \ref{paratab} can be directly compared, since
for $E=0$, one has $\Delta m /A^2 = m_2^{AA} $, cf.~(\ref{quadform}).}
listed in the last line of Table~\ref{slopetab}, providing an independent
measurement corroborating the interpretation of the data advanced in
section \ref{a3sec}.

\begin{table}[ht]
\begin{center}
\begin{tabular}{|c||c|c|c|c|}
\hline
& (a) & (b) & (c) & (d) \\
\hline \hline
\parbox{2.5cm}{$-\Delta m /(a^3 E^2 )$ \\ \mbox{\hspace{-0.1cm} }
(extremum) } & $-0.034(6)$ & $-0.049(10)$ & $-0.052(24)$ & $-0.076(63)$ \\
\hline
$-m_2^{AA} /a$ & $0.0027(9)$ & $0.0031(9)$ & $0.0018(16)$ & $0.0022(16)$ \\
\hline
\end{tabular}
\end{center}
\caption{(Negative) mass shifts at the extrema of the parabolas defined
by the data in Table~\ref{slopetab} through the form (\ref{parafit}),
as well as the coefficient $m_2^{AA} $ characterizing the curvatures of
the parabolas.}
\label{paratab}
\end{table}

The central result of this work, however, is the value of the electric
polarizability of the neutron, the full value of which is obtained by
multiplying the mass shift quoted in column (c) of Table~\ref{paratab}
by a factor of 2, cf.~(\ref{quadform}),(\ref{alphaEE}). In physical
units, obtained by inserting $a=0.124\, \mbox{fm} $, one has
\begin{equation}
\alpha = -2\Delta m/E^2 = (-2.0\pm 0.9)\cdot 10^{-4} \, \mbox{fm}^3 \ .
\label{alphafin}
\end{equation}
Note that the additional smearing contributions entering the
result in column (d) indeed do not significantly alter this result.
The result (\ref{alphafin}) is corroborated by the measurement using
the external field (\ref{a0def2}), quoted in the fourth line of
Table~\ref{slopetab}; translated into physical units, that measurement
would imply a polarizability of
\begin{equation}
\alpha = (-0.7\pm 1.6)\cdot 10^{-4} \, \mbox{fm}^3 \ .
\label{alphaside}
\end{equation}
As discussed in section \ref{a0sec}, the result (\ref{alphaside}) contains
systematic uncertainties (not included in the quoted statistical error)
stemming from the fact that the mass shift measured in this case is
contaminated by the quantum mechanical zero-point motion of the neutron
and distortions of its internal wave function due to superfluous spikes
in the external electric field. Thus, the result (\ref{alphafin}) is
expected to be more trustworthy than the result (\ref{alphaside}).
Nevertheless, the difference between the two measurements does not turn
out to be significant; the aforementioned contaminations do not appear
to represent appreciable effects.

Compared to the experimental value reported by the Particle Data Group
\cite{pdgroup}
\begin{equation}
\alpha = (11.6\pm 1.5) \cdot 10^{-4} \, \mbox{fm}^3
\label{pdg}
\end{equation}
the result (\ref{alphafin}) suggests a strong variation of the electric
polarizability of the neutron with the pion mass. Indeed, Chiral Effective
Theory calls for such a variation \cite{detmold,vbern,kambor,hgrie,bira},
dominated by a $1/m_{\pi } $ dependence at low pion masses. In the
``Small Scale Expansion'' approach \cite{kambor,hgrie}, which
systematically extends leading-one-loop Heavy Baryon Chiral
Perturbation Theory by including explicit $\Delta $ degrees of
freedom, the electric polarizability of the neutron decreases
by an order of magnitude as one varies the pion mass from the
physical point up to around $400\, \mbox{MeV} $. Qualitatively,
a change of sign of the polarizability at even higher pion masses, as
implied by (\ref{alphafin}), does not seem implausible\footnote{Of
course, for very large masses, i.e., in the nonrelativistic limit,
the polarizability cannot be negative due to the general properties of
second-order perturbation theory.}, although it should be stressed that
a pion mass of $759\, \mbox{MeV} $, corresponding to the dynamical quark
ensemble used in the present work, is certainly far beyond the regime in
which Chiral Effective Theory can be applied reliably. Lattice calculations
at lower pion masses are needed in order to achieve a quantitative
connection with Chiral Effective Theory.

On the other hand, the result obtained in the present work at first sight
appears to be at odds with previous lattice measurements \cite{wile,wilepap}.
Those studies yield a neutron electric polarizability which is
consistent with (\ref{pdg}) over a wide range of (valence) quark
masses\footnote{The studies reported in \cite{wile,wilepap} employ the
quenched approximation.}, including the quark mass used in the present
work. The question arises how such a weak variation with the quark mass
can be reconciled with the result obtained in the present work (and
also with the expectation coming from Chiral Effective Theory).

Apart from the use of the quenched approximation, the main differences
between \cite{wile,wilepap} and the present treatment are that, on the one
hand, \cite{wile,wilepap} work with an external field corresponding to a
particular value of $t_0 $ in (\ref{a3def2}); on the other hand, this
external field is introduced into the lattice link variables in linearized
form, i.e., the vertex insertion $M_2 $, cf.~(\ref{se2}), is not included.
As discussed further above, the insertion $M_2 $ provides contact terms
which renormalize propagators with two $M_1 $ insertions, leading to
substantial cancellations; however, most importantly, the measured mass
shift depends sensitively on the parameter $t_0 $ characterizing the
external field. It is instructive to reevaluate the data gathered in the
present work such that diagrams generated by $M_2 $ insertions are excluded,
and at a value of $t_0 $ corresponding to the one used in
\cite{wile,wilepap}. Two measurements are provided by \cite{wile,wilepap}.
One uses Wilson fermions with a distance of $1.7\, \mbox{fm} $ between
$t_0 $ and the mass shift measurement, yielding a 
polarizability\footnote{These estimates were obtained by linearly
interpolating results quoted in \cite{wilepap} in $m_{\pi } $.}
of $\alpha = (9.8\pm 1.2)\cdot 10^{-4} \, \mbox{fm}^3 $ for pion masses
comparable to the one at which the present work was performed; the other
uses clover fermions with a distance of $1.53\, \mbox{fm} $ between $t_0 $
and the mass shift measurement, yielding a polarizability of
$\alpha = (13.9\pm 0.8)\cdot 10^{-4} \, \mbox{fm}^3 $ at comparable pion
masses. Taking into account that the mass shift measurement in the
present work is centered around $t=7a=0.87\, \mbox{fm} $, the values
of $t_0 $ corresponding to the two aforementioned cases are $t_0 = -6.7a$
and $t_0 = -5.3a$, respectively. Constructing the parabola (\ref{parafit})
defined by excluding diagrams generated by $M_2 $ insertions from the
set of diagrams comprising case (c) above, and evaluating it at those values
of $t_0 $ yields
\begin{eqnarray}
\alpha (t_0 = -6.7a) &=& (20\pm 11)\cdot 10^{-4} \, \mbox{fm}^3 \\
\alpha (t_0 = -5.3a) &=& (15\pm 8)\cdot 10^{-4} \, \mbox{fm}^3
\end{eqnarray}
which, particularly in the latter case, is in quite good
agreement\footnote{It should be noted that these comparisons depend
sensitively on the determinations of the lattice spacings in the
different calculations.} with the results of \cite{wile,wilepap}. Thus,
at the level of the raw numerical measurement, the present work in
fact corroborates the results obtained in \cite{wile,wilepap}; at the
same time, it is now clear that such a measurement at a single fixed
$t_0 $ in general contains two separate effects, i.e., the electric
polarizability itself and the effect of subjecting the neutron to
a constant external field. To disentangle the former from the
latter using such measurements, it is necessary to combine the data
obtained using a variety of $t_0 $. This suggests that it would be
worthwhile to supplement the measurements already performed in
\cite{wile,wilepap} by further analogous measurements at other values
of $t_0 $ in order to obtain a more comprehensive picture of the physical
effects engendered by the introduction of the external field and thus be
able to isolate the different effects from one another.

\section{Summary and outlook}
The investigation reported here represents a first exploration of the
neutron electric polarizability in the context of lattice QCD with
dynamical quarks. Its main thrust lay in clarifying conceptual questions
within the framework of the background field method and assessing the
feasibility of numerical computations on that basis, using a $SU(3)$
flavor-symmetric ensemble as a test case.

Two central issues needed to be addressed to arrive at a cogent
calculational scheme. On the one hand, the presence of dynamical quarks
dictates the use of four-point function methods, introducing, in particular,
the need to evaluate disconnected diagrams. These contributions, which
were included in the numerical calculations carried out in this work via
stochastic estimation, significantly increase the computational expense
of the measurement. Nevertheless, the feasibility of carrying out such
measurements was demonstrated for the $SU(3)$ flavor-symmetric case;
the cost of progress towards lighter quark masses does not seem prohibitive,
but such an endeavor will require a, by current standards, significant
commitment of computational resources.

On the other hand, a strong emphasis was placed within the present work on
the physical consequences of shifting the external electromagnetic field
by a constant on a finite lattice. While such shifts merely
correspond to gauge transformations in infinite space, on a finite lattice,
they influence the physical spectrum and thus mask the mass shift due to
the electric polarizability itself. On lattices of a practical size,
this effect has a dominating influence on the neutron mass shift, from
which one aims to extract the electrical polarizability. To disentangle
the two effects, measurements using a variety of external fields which
are shifted with respect to one another are necessary (further impacting
computational cost). It should be noted that this issue also affects
investigations carried out in the quenched approximation, such as reported
in \cite{wile,wilepap}. The present investigation, complementing that
effort, suggests additional measurements to supplement the ones
already carried out, in order to gain a comprehensive picture of the
effects playing a role. It is hoped that the results obtained here will
provide motivation and useful input for an expanded measurement program
in this direction.

Looking forward, besides the obvious need to progress towards lighter
quark masses, it would be interesting to study other hadrons, especially
with a view towards measuring polarizability combinations in which
disconnected diagrams at least partially cancel. Such combinations could
be calculated with higher accuracy at lower cost. However, a potential
obstacle to this which should be kept in mind is the following: Typically,
hadrons of differing electric charge would be involved, and, a priori,
it is not clear that measurements using the same external electromagnetic
field are appropriate in each case for the purpose of isolating the
electric polarizability. On the other hand, results obtained in different
external fields cannot be combined straightforwardly to cancel disconnected
contributions.

One possibility of avoiding such difficulties lies in using alternate
methods of accessing polarizabilities, e.g., via density-density
correlation functions \cite{grandypol}. That approach would circumvent
the necessity of explicitly introducing an external electromagnetic
field. Density-density correlation functions at unequal times can be used
to extract hadron polarizabilities, specifically by measuring the second
moment (with respect to spatial separation) of the correlation function
for a range of relative times and integrating over the latter. When
calculating polarizabilities of hadrons in this manner, at least partial
cancellations of disconnected diagrams can be achieved straightforwardly
by forming the proper isovector combinations. Care must be taken to
restrict the hadron momentum to the nonrelativistic regime, in order
to exclude relativistic effects which complicate the interpretation
of the density-density correlation function and the extraction of the
polarizability. Also, density-density correlation functions generally
fall off less rapidly than standard hadron wave functions (a doubling
of the extent being typical); this has motivated the development of periodic
image correction methods \cite{grandypol} which are expected to prove
helpful in this context.

\section*{Acknowledgments}
The author is especially grateful to J.~Negele, K.~Orginos and D.~Renner
for sharing their expertise in numerous discussions. This investigation
furthermore benefited from helpful exchanges with R.~Brower, M.~Burkardt,
W.~Detmold, R.~Edwards, H.~Grie\ss hammer, J.~Osborn, D.~Toussaint and
W.~Wilcox, as well as from enlightening comments provided by D.~B.~Kaplan
and M.~Savage. Also, this work would not have been possible without the
dynamical quark configurations made available by members of the MILC
Collaboration. It is a pleasure to acknowledge the use of computer
resources provided by the U.S.~DOE through the USQCD project at
Jefferson Lab, and support by the U.S.~DOE under grant number
DE-FG03-95ER40965.
\vspace{0.6cm}

\noindent
{\em Note added:} The author is grateful to D.~Toussaint for pointing out
ref.~\cite{shintani}, which in an appendix also discusses the electric
polarizability of the neutron, and for further exchanges thereon. One
particular point emphasized in \cite{shintani} is that, to properly
represent the effect of a classical external electric field, the electric
field as introduced in this work and also in \cite{fiebpol,wile,wilepap}
should be analytically continued to imaginary values. This further step,
which was not carried out in \cite{fiebpol,wile,wilepap} and in the
present treatment, implies that the results (\ref{alphafin}) and
(\ref{alphaside}) receive an additional overall minus sign. In view of
the small magnitude of (\ref{alphafin}) and (\ref{alphaside}), this does
not decisively impact the further conclusions drawn in the present work.
However, it will need to be taken into account, and revisited in more
detail, in work going forward aiming at progress towards lighter pion
masses.

\end{document}